\documentclass[twocolumn,aps,showpacs,
amsfonts,amsmath,amssymb,
superscriptaddress,floatfix]
{revtex4}
\usepackage[dvips]{graphicx}

\newcommand{\ket}[1]{\left | #1 \right \rangle}
\newcommand{\bra}[1]{\left \langle #1 \right |}
\newcommand{\amp}[2]{\left \langle #1 | #2 \right \rangle}
\newcommand{\proj}[1]{\ket{#1} \! \bra{#1}}
\newcommand{\tr}{{\rm \, Tr }\, }
\newcommand{\be}{\begin{equation}}
\newcommand{\ee}{\end{equation}}
\newcommand{\affA}{%
     Communications Research Laboratory,
     Koganei, Tokyo 184-8795, Japan}
\newcommand{\affB}{%
     CREST, Japan Science and Technology Corporation, 
     Shibuya, Tokyo 150-0002, Japan}
\begin{document}

\title{Implementation of generalized quantum measurements: 
superadditive quantum coding, accessible information extraction, and 
classical capacity limit}

\author{Masahiro Takeoka}
\author{Mikio Fujiwara}
\author{Jun Mizuno}
\author{Masahide Sasaki}
\affiliation{\affA}
\affiliation{\affB}

\begin{abstract}

Quantum information theory predicts that
when the transmission resource is doubled in quantum channels,
the amount of information transmitted can be increased more than twice
by quantum channel coding technique, 
whereas the increase is at most twice in classical information theory.
This remarkable feature, the superadditive  quantum coding gain, 
can be implemented by appropriate choices of code words and 
corresponding quantum decoding which requires 
a collective quantum measurement. 
Recently, the first experimental demonstration was reported 
[Phys.~Rev.~Lett.~{\bf 90}, 167906 (2003)]. 
The purpose of this paper is to describe our experiment in detail. 
Particularly, a design strategy of quantum collective decoding in 
physical quantum circuits is emphasized. 
We also address the practical implication of 
the gain on communication performance 
by introducing the {\it quantum-classical hybrid coding} scheme. 
We show how the superadditive quantum coding gain, 
even in a small code length, 
can boost the communication performance 
of conventional coding technique. 

\end{abstract}

\pacs{03.67.-a, 03.65.Ta, 89.70.+c}
%
\date{\today}
\maketitle

\section{Introduction}\label{introduction}

It is a fundamental problem in information science 
what is the most efficient way of transmitting information 
with a minimum of transmission resources. 
The amount of information transmissible 
through a communications channel is determined 
by the noise characteristics of the channel 
and 
by the quantities of available transmission resources. 
In classical communication theory
\cite{Shannon48,Gallager_book,CoverThomas_book}, 
the amount of transmissible information 
can be increased twice at most 
when the transmission resource 
(e.g. the code length, the signal power, the bandwidth) 
is doubled, 
for the fixed noise characteristics of the channel. 
In quantum communication theory, 
however, 
this is not true in general, 
that is, 
the amount of information transmitted can be increased 
even more than twice.
This feature is called 
the {\it superadditivity} of the capacity of quantum channel 
\cite{Holevo79_QuantCap,PeresWootters91,Sasaki97_SupAdd,
Sasaki98_SupAdd,Buck00_SupAdd,Usuda02_SupAdd}.

The superadditivity becomes essential 
in any transmission of signals at the quantum level 
where 
ambiguity among signals is a matter of non-commutativity 
of the density matrices, 
i.e. $\hat\rho_0\hat\rho_1 \ne \hat\rho_1\hat\rho_0$, 
rather than any classical noises such as thermal noise. 
One typical example is 
deep space optical communications. 
Inter-satellite optical link is expected to achieve 
a high transmission rate 
that cannot be achieved by the radio- or micro-wave links,  
e.g. 
to realize data transmission from a space-telescope with billion pixels 
or 
real time communications over the planets. 
In a deep space optical link, 
the sender prepares coherent state signals 
with as large amplitude as possible 
allowed by a limited power supply. 
Such signals are, however, extremely weakened at the receiving end, 
typically less than a few photons per pulse, 
due to the beam divergence and energy loss. 
Since 
the energy quantum of carriers is greater than 
that of thermal noise 
in optical domain, 
i.e. $\hbar\omega>k_\mathrm{B}T$, 
physical states of carriers can be described by pure quantum states 
in good approximation. 
For example, 
the binary phase shift keyed signals 
to convey classical letters 0 and 1, respectively, 
are represented by the coherent states 
$\hat\rho_0=\proj{\alpha}$ 
and $\hat\rho_1=\proj{-\alpha}$, respectively. 
For weak coherent pulses, the state overlap 
$\langle\alpha\vert \! - \! \alpha\rangle$ becomes non-negligible, i.e. 
$\hat\rho_0$ and $\hat\rho_1$ are noncommuting. 
According to the uncertainty principle, 
noncommuting density matrices can never be distinguished perfectly. 
This imposes an inevitable error 
in signal detection even in an ideal communications system 
\cite{Helstrom_QDET}. 
Actually, when $|\alpha|^2 \le 3$, 
$\hat{\rho}_0$ and $\hat{\rho}_1$ 
can not be distinguished 
at a bit error rate less than $10^{-6}$, 
which is a typical error free criterion in deep space communications.

Historically, 
an extension of communication theory into quantum domain 
including this aspect of ambiguity 
has been explored since 1960's 
\cite{Gordon62,Gordon64_bound,Lebedev66,Holevo73_bound}. 
In 1973, Holevo derived the quantity that bounds 
the upper limit of the capacity of a quantum communications channel 
\cite{Holevo73_bound}. 
It was recently shown that 
this so-called Holevo bound is 
an achievable rate, that is, 
the exact expression of the capacity 
\cite{Hausladen96_capacity,Schumacher97_capacity,Holevo98_capacity}. 
Classical communication theory 
\cite{Shannon48,Gallager_book,CoverThomas_book} 
describes the special case where the signals 
are given by commuting density matrices. 
The distinctive characteristics of quantum theory of capacity 
is the great emphasis on quantum decoding process 
to extract information from 
block sequences of noncommuting density matrices. 
The essence of the optimal decoding is the use of a process of entangling 
letter states constituting code words prior to measurement 
to enhance the distinguishability of signals. 
Such a process is 
a quantum computation on code word states. 
This so-called quantum collective decoding is a new aspect, 
not found in conventional coding techniques, 
and leads to a larger capacity. 
This is called the {\it superadditive quantum coding gain} (SQCG)  
\cite{Holevo79_QuantCap,PeresWootters91,Sasaki97_SupAdd,Sasaki98_SupAdd,
Buck00_SupAdd,Usuda02_SupAdd} in a quantum channel 
since the length $n$ quantum coding makes capacity  
more than $n$ times larger from the capacity 
achievable only by conventional coding. 
Taking it into account, 
the capacity is defined as the maximum rate of 
the mutual information for a (quantum) code of length $n$ 
divided by the length $n$ 
in the limit of $n\rightarrow\infty$ 
for asymptotic error free transmission.

The theory of capacity, 
however, 
generally gives no guidance on 
how to construct codes that approach the capacity. 
A practical problem is then 
to find good codes to attain a large SQCG 
in a small block length. 
This must be an important issue 
in any communications and information-processing systems 
when they work at the quantum level, 
which is expected in a few decades 
considering recent exponential growth 
of info-communication demands. 
However, 
little attention has been paid to this topic so far. 
Only several coding schemes have been proposed 
to exhibit SQCG 
\cite{PeresWootters91,Sasaki97_SupAdd,Sasaki98_SupAdd,
Buck00_SupAdd,Usuda02_SupAdd}
and 
the first experimental demonstration has recently been reported 
by the authors 
\cite{FujiwaraTakeokaMizunoSasaki03_ExpSupAdd}. 
The purpose of the present paper is to give detailed 
information that was abbreviated or omitted 
in our previous letter. 
An attention is particularly paid to describe the strategy how to 
implement the quantum measurements used in our experiments 
by logical and physical quantum circuits. 
We also describe the detailed discussion of 
the implications of SQCG in small code length 
on practical communication performances.

The paper is organized as follows.
In Sec.~\ref{capacities},
we remind readers of
several capacities of quantum channels studied to date,
and explain our scenario. 
In Sec.~\ref{SupAddCodingGain}, 
the basic notion for capacity theorem and 
SQCG are briefly explained. 
In Sec.~\ref{Model}, 
we discuss how we designed logical and physical quantum circuits 
for our SQCG experiment, 
which was omitted in our previous letter. 
Section \ref{Implementation} describes in detail 
our experiment of SQCG reported in 
Ref.~\cite{FujiwaraTakeokaMizunoSasaki03_ExpSupAdd}. 
We also show the experimental results 
about the separable quantum measurements attaining 
the single--shot capacity and 
the accessible information for comparison. 
In Sec.~\ref{QCHC}, 
we discuss how SQCG, 
even the small gain demonstrated in length two coding, 
can boost a communication performance 
attained by conventional coding technique. 
The idea is based on {\it quantum-classical hybrid coding} (QCHC)
which was briefly mentioned in our previous letter 
\cite{FujiwaraTakeokaMizunoSasaki03_ExpSupAdd}. 
Theoretical details on the methodology of QCHC are given. 
Section \ref{concluding} is for concluding remarks.

\section{capacities for quantum channels}
\label{capacities}

Since Shannon's capacity theory was extended into 
generic quantum states in Refs.~\cite{Hausladen96_capacity,
Schumacher97_capacity,Holevo98_capacity}, 
the capacity theory is further extended to include 
new auxiliary resources of entangled particles, 
new quantum protocols, 
and 
a new object to be transmitted, 
i.e. intact quantum state
\cite{Q_cap}. 
The notion of the capacity for quantum channels 
is now classified into two categories; 
1) the classical capacity 
for transmitting conventional (classical) alphabet, 
and 
2) the quantum capacity 
for transmitting quantum alphabet 
(unknown quantum states). 
For both categories 
entanglement assisted protocols may be considered, 
namely superdense coding 
\cite{BennettWiesner92} 
and quantum teleportation 
\cite{Bennett92}, 
respectively. 
Our concern is the first category, i.e. 
the {\it classical} capacity.

Depending on whether additional entanglement resources are 
brought into play or not, 
the classical capacity is classified 
into two kinds, 
namely, 
the entanglement-assisted capacity $C_{\mathrm E}$ 
and 
the ordinary capacity $C$. 
The former is defined for a quantum channel with the help of 
unlimited prior entanglement sharing between the sender and the receiver 
\cite{BennettShorSmolinThapliyal02}.
The latter is defined for a quantum channel with the help of 
any allowed quantum operations at the sender (quantum encoding) 
and the receiver (quantum decoding), 
but {\it without} any prior source sharing. 
Both schemes assume multiple uses of the channel, that is, coding, 
and the capacity is defined as the maximum amount of 
transmissible information per channel use. 
It should be noted that 
shared entanglement is not regarded as the transmission resources 
in the definition of the entanglement-assisted capacity $C_{\mathrm E}$. 
This type of classification can simplify the study of several 
distinct capacities and their relation 
including the quantum capacities
\cite{BennettShorSmolinThapliyal02}.

From a practical point of view, 
on the other hand, 
it is not realistic to expect the assistance of 
{\it unlimited} external resources. 
To predict the highest transmission rate in realistic situations, 
all the physical entities used for transmission, 
such as the shared entanglement, must be 
included in the elements constituting a communications channel. 
In this situation, 
the ordinary capacity $C$ is appropriate 
to evaluate the communication performance of the channel 
since it imposes the power constraint condition of 
the {\it total} physical resources.

Keeping such backgrounds in mind, 
we restrict our discussion to the following protocol. 
The sender transmits classical alphabet 
in classically encoded format, 
i.e. 
in separable tensor product states (code word states) 
made of a given set of {\it letter states}, 
and 
these code word states are still separable at the receiving end. 
No prior entanglement is shared between the sender and the receiver. 
The receiver may apply any quantum operations 
to the received code word states. 
In fact, it is known that effective quantum coding is that 
the receiver entangles 
the letter states prior to detection. 
This is called quantum decoding and contributes to SQCG 
which is never observed in any classical coding. 
Such scenario is within the framework of the ordinary capacity $C$, and 
exactly the case that the capacity theories of Refs.  
\cite{Hausladen96_capacity,
Schumacher97_capacity,
Holevo98_capacity} 
concern. 
The reasons for choosing such protocol and 
for excluding quantum encoding and 
prior entanglement sharing 
are 
1) concrete quantum coding schemes are known only for such protocol 
at present, and 
2) it fits better to practical motivations 
introduced in Sec.~\ref{introduction}
since it does not necessarily require 
a transmission of nonclassical state signals. 
The main task in this paper is, therefore, 
the demonstration of the quantum decoding process.

\section{superadditive coding gain}
\label{SupAddCodingGain}

A practical mean for effective communications is coding, 
that is, representing alphabet 
by sequences of simple letters 
such as $\{0,1\}$. 
Alphabets to be transmitted are represented by code words 
which are sequences of a given set of letters 
$\{ x_0, ..., x_{L-1} \}$ 
such as 
the binary set $\{0,1\}$. 
The transmitter modulates a signal carrier into one of $L$ states 
$\{ \hat\rho_0, ..., \hat\rho_{L-1} \}$ 
according to the input letter. 
If the letter states $\{ \hat\rho_0, ..., \hat\rho_{L-1} \}$ 
appear as orthogonal states at the receiving end, 
then they can be distinguished perfectly 
and $\log_2 L$ bits of information, 
which is the maximum Shannon entropy of the set 
$\{ x_0, ..., x_{L-1} \}$, 
can be faithfully retrieved per letter.  
This is, however, not the case in general. 
A channel is usually subject to 
various types of noise disturbances. 
In order to transmit information reliably through a channel 
with finite errors, 
one must introduce some redundancy in code word representation 
prior to transmission 
so as to allow the correction of errors at the receiving side. 
This entails adding some redundant letters to the code words 
and hence increases their length.
This is \textit{channel coding}.

First, the source encoder converts the original message into 
a sequence of the letters in the given set $\{ x_0, ..., x_{L-1} \}$, 
and then the channel encoder divides it into blocks of length $k$ 
(\emph{message blocks}).
Each block is supplemented by an additional block
(\emph{redundant block}) of $n{-}k$ ($n{>}k$) letters
to compose a channel code word $\{\mathbf{x}_i\}$\,:
\begin{eqnarray}
\mathbf{x}_i &=&
\overbrace{x_{1}^{(i)} x_{2}^{(i)}\cdots x_{k}^{(i)}}%
   ^{\displaystyle\text{\mathstrut message block}}\,
\overbrace{x_{k+1}^{(i)} x_{k+2}^{(i)} \cdots x_{n}^{(i)}}%
   ^{\displaystyle\text{\mathstrut redundant block}}
\\
&&\text{(for  $i=1,2,\ldots,L^k$)}
\,.
\nonumber
\end{eqnarray}
Note that although there are $L^n$ possible sequences
of length~$n$ in total,
only part of them, i.e.\ $L^k$ sequences,
are used as code words. 
This redundancy,
together with appropriate encoding and decoding, 
allows us to recover possible errors in transmission. 
The amount of information conveyed by the above code words is 
$K=k\log_2 L$ bits. 
The transmission rate is then defined by 
$R={K/n}=(k/n)\log_2 L$ 
bits/letter. 
For a channel with a capacity $C$ bits/letter, 
it is possible
\cite{Shannon48,Gallager_book,CoverThomas_book} 
within the rate $R=K/n < C$ to reproduce the $K$ bits of messages 
with an error probability as small as desired by appropriate encoding 
and decoding in the limit $n\rightarrow\infty$.

A mathematical model of a channel is specified 
by a set of possible outputs $\{ y \}$ from the channel 
and 
a channel matrix in which each matrix element is given by 
the conditional probability $P(y\vert x)$ of having $y$ 
given the input $x$. 
Each input letter $x$ is used with a priori probability $P(x)$. 
The probability of having $y$ is then given by 
\begin{equation}
P(y)\equiv\sum_{x} P(y \vert x)P(x). 
\label{P(y)}
\end{equation}
To define the capacity, 
Shannon introduced the mutual information
\cite{Shannon48}. 
This is defined between the input variable $X=\{ x ; P(x) \}$ 
and 
the output variable $Y=\{ y ; P(y) \}$ as  
\begin{equation}
I(X:Y) = 
\sum_{x} P(x)
\sum_{y} P(y\vert x)
\log \biggl[ { P(y\vert x) \over
                  {\displaystyle\mathop{\textstyle\sum}_{x'}
                         P(x')P(y\vert x')}}
     \biggr]
\,.
\label{IXY}
\end{equation}

In classical information theory, 
one considers coding for a given and fixed channel model 
$\{P(y\vert x)\}$. 
The decoding error of code words 
$\{ {\mathbf x}_1,...,{\mathbf x}_{L^k} \}$ 
can be calculated based on the probability distributions 
$\{P(x)\}$ and $\{P(y\vert x)\}$. 
The capacity (for a memoryless channel) 
is defined as the maximum mutual information with 
respect to the prior distribution of the letters $P(x)$, 
\begin{equation}
C=\max_{\{P(x)\}} I(X:Y) \, . 
\label{C_def}
\end{equation}

In the quantum context, however, 
only the input variable $X$ 
and 
the corresponding set of quantum states 
at the receiver's hand denoted as 
$\{ \hat\rho_x \}$ are given. 
The output variable $Y$ is to be sought for the best quantum 
measurement. 
A quantum measurement process can mathematically be described 
by a set of non-negative Hermitian operators $\{\hat\Pi_y\}$ 
satisfying the probability conservation relation 
$\sum_y \hat\Pi_y=\hat I$, 
so-called the positive operator valued measure (POVM). 
The channel matrix is then given by 
\begin{equation}
P(y\vert x)\equiv \tr \left( \hat\Pi_y \hat\rho_x \right), 
\label{P(y|x)} 
\end{equation} 
and now one can define the maximum extractable information 
\begin{equation}
I_{\rm Acc} = \max_{\{\hat\Pi_y\}} I(X:Y), 
\label{I_Acc}
\end{equation}
which is called the accessible information. 
More generally, 
the quantity further maximized with respect to the prior probability 
\begin{equation}
C_1=\max_{\{P(x)\}}\max_{\{\hat\Pi_y\}} I(X:Y), 
\label{C1_def}
\end{equation}
specifies the classical limit of the capacity 
when the given initial channel $\{P(y\vert x)\}$ 
is used with classical channel coding 
\cite{FujiwaraNagaoka98}. 
It is this quantity that limits the performance of all modern 
communications systems.  
This is, however, not the ultimate capacity allowed by quantum mechanics.

The code words $\{ {\mathbf x} \}$ are now conveyed 
by the quantum states in a tensor product of the letter states, 
$\hat\Psi_{\mathbf x}
=\hat\rho_{x_1}\otimes\cdots\otimes\hat\rho_{x_n}$. 
To decode them, one may design the best quantum measurement 
allowed by quantum mechanics. 
This is described by the POVM $\{ \hat\Pi_{\mathbf y} \}$ 
on the extended space 
where $\{ {\mathbf y} \}$ are decoded code words. 
The channel matrix for this extended channel is given by 
\begin{equation}
P({\mathbf y}\vert{\mathbf x})
\equiv 
\tr \left( \hat\Pi_{\mathbf y} \hat\Psi_{\mathbf x} \right). 
\label{extended_channel_matrix} 
\end{equation}
One may then define the mutual information 
for this extended channel 
by 
\begin{equation}
I(X^n:Y^n) = 
\sum_{{\mathbf x}} P({\mathbf x})
\sum_{{\mathbf y}} P({\mathbf y}\vert {\mathbf x})
\log \biggl[ 
     { 
       P({\mathbf y}\vert {\mathbf x}) 
       \over
       {\displaystyle\mathop{\textstyle\sum}_{{\mathbf x}'}
        P({\mathbf x}')P({\mathbf y}\vert {\mathbf x}')} 
     }
     \biggr]
\,.
\label{IXY_extended}
\end{equation}
Further one can define the quantity
\begin{equation}
C_n=\max_{\{P({\mathbf x})\}}\max_{\{\hat\Pi_{\mathbf y}\}} 
I(X^n:Y^n), 
\label{Cn_def}
\end{equation}
which we refer to the capacity of order $n$. 
The superadditivity of quantum channel is then expressed as 
\begin{equation}
C_n>n C_1. 
\label{Cn>nC1}
\end{equation} 
The capacity of quantum channel 
as the maximum rate of error free transmission is 
defined by 
\begin{equation}
C=\lim_{n\rightarrow\infty}\frac{C_n}{n}. 
\label{C_Q_def}
\end{equation}

The property of Eq.~(\ref{Cn>nC1}) 
was first predicted by Holevo based on 
the random coding technique 
\cite{Holevo79_QuantCap}. 
Peres and Wootters conjectured that 
\begin{equation}
I(X^2:Y^2)>2 \max_{\{\hat\Pi_y\}} I(X:Y),  
\label{PeresWootters_conjecture}
\end{equation}
by using the ternary symmetric states of qubit
\cite{PeresWootters91}. 
In these works, 
the importance of using quantum collective measurement on a block 
sequence of code word state was emphasized. 
The first rigorous example of the superadditivity was given 
by Sasaki {\it et al.} for the binary pure letter states 
\cite{Sasaki97_SupAdd}, where the quantum channel showing 
\begin{equation}
I(X^3:Y^3)>3 C_1,  
\label{Sasaki_coding97}
\end{equation}
was explicitly demonstrated. 
Since then several examples of quantum code construction with 
the superadditivity were clarified 
\cite{Sasaki97_SupAdd,Sasaki98_SupAdd,Buck00_SupAdd,Usuda02_SupAdd}.

The important observation of the superadditivity is 
that the property 
\begin{equation}
P({\mathbf y}\vert{\mathbf x})
\ne
P(y_1\vert x_1)\cdots P(y_n\vert x_n), 
\label{memory_effect} 
\end{equation}
generally holds 
when an appropriate collective POVM is chosen. 
This is a kind of memory effect of the extended channel. 
When the measurement is made by 
projection onto separable bases, 
such a memory effect never takes place. 
Projection onto appropriate entangled bases 
induces quantum interferences among the code word states 
to reduce the ambiguity among the signals. 
The memory effect is a direct consequence of 
this quantum interference of block codes, that is, 
exactly the effect of the entanglement. 
Realization of such a quantum collective decoding generally 
requires quantum computation to entangle the letter states 
\cite{Sasaki98_realization}.

\section{Model for proof-of-principle demonstration:
qubit trine}
\label{Model}

Quantum collective decoding on quantum particles, 
that is, entangling quantum particles, is 
something very hard to realize at present even for two particles. 
In addition, 
the gains predicted for short length codes are very small. 
Here we consider how one can demonstrate the principle of 
quantum collective decoding. 
We deal with the noiseless channel model 
in which only the noncommutativity of the signals causes 
the transmission error.

The simplest set of letters is the binary set of pure states 
$\{ \ket{\psi_0},\ket{\psi_1} \}$ 
where 
the overlap between the letters is 
$\langle \psi_0 | \psi_1 \rangle = \kappa$. 
Only for this set, 
the classical capacity limit $C_1$ 
is known with rigorous mathematical proof 
\cite{Levitin95_QCM94,Osaki2000_QCM98_C1,Davies78}. 
The very first step is the length two coding. 
We have four possible tensor product sequences 
$\{ 
\ket{\psi_0}\ket{\psi_0},
\ket{\psi_0}\ket{\psi_1},
\ket{\psi_1}\ket{\psi_0},
\ket{\psi_1}\ket{\psi_1} 
\}$.
Buck {\it et al.} 
\cite{Buck00_SupAdd} 
showed that, by choosing three of them as code word states, 
the channel exhibits the superadditivity 
depending on the overlap $\kappa$. 
Unfortunately, 
the predicted maximum SQCG was only 
$I_2/2-C_1=5.2 \times 10^{-4}$ bits 
and 
it seems too small to be observed experimentally. 
In the length three coding \cite{Sasaki97_SupAdd}, 
the four code words were picked up from eight possible sequences 
so that the Hamming distance between each code word 
is equal ($d=2$) 
to show the gain described in Eq.~(\ref{Sasaki_coding97}). 
In this case, 
the maximum gain was predicted to be $9 \times 10^{-3}$ bits. 
Although the gain is bigger than that in the length two coding, 
entangling three qubits for the quantum collective decoding 
requires more than ten steps of quantum gating, 
which seems to be difficult to realize.

\begin{figure} 
\begin{center} 
\includegraphics[width=0.45\textwidth]{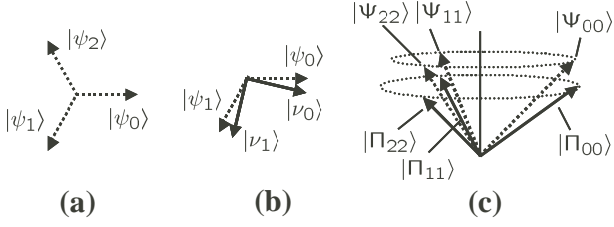} 
\end{center} 
\caption{\label{fig:ternary_letter_set} 
Geometrical representation 
of several sets of quantum state vectors 
and measurement vectors. 
(a) The ternary symmetric letter states (qubit trine). 
(b) The letter (dotted arrows) 
and the measurement (solid arrows) state vectors 
to attain the $C_1$.
(c) The code word (dotted arrows) 
and decoding (solid arrows) state vectors 
represented in a real three dimensional space. 
}
\end{figure}

Therefore, we consider the second simplest case, 
the qubit trine signals. 
The qubit trine consists of 
the ternary symmetric letter states of a qubit 
$\{ \ket{\psi_0},\ket{\psi_1},\ket{\psi_2} \}$. 
It is this model 
which we use to demonstrate our experimental steps 
toward SQCG in the next section. 
It should be noted that 
the dimensionality (2--dim.~) is essential here. 
If the ternary states are defined 
in a higher dimensional space than three, 
such as the {\it lifted} trine, 
we do not know the exact value of $C_1$. 
In addition, according to our numerical studies, 
SQCG in terms of 
the mutual information appears smaller compared to 
that in the qubit trine case. 
The qubit trine is defined by the letter state set 
$\{ |\psi_0\rangle, |\psi_1\rangle, |\psi_2\rangle \}$ 
with
\begin{subequations}
\label{ternary_state}
\begin{eqnarray}
| \psi_0 \rangle 
& = & | 0 \rangle,
\\
| \psi_1 \rangle 
& = & - \frac{1}{2} | 0 \rangle 
      - \frac{\sqrt{3}}{2} | 1 \rangle,
\\
| \psi_2 \rangle 
& = & - \frac{1}{2} | 0 \rangle 
      + \frac{\sqrt{3}}{2} | 1 \rangle,
\end{eqnarray}
\end{subequations}
where 
$\{ |0\rangle, |1\rangle \}$ is the orthonormal basis set. 
They are represented in 
Fig.~\ref{fig:ternary_letter_set}(a).

The accessible information, defined by Eq.~(\ref{I_Acc}), 
for this set with equal prior probabilities 
is found to be $I_{\rm Acc}=0.5850$ bits 
with a rigorous proof 
\cite{SasakiBarnettJozsa99}. 
The optimal measurement strategy is described by 
the nonorthogonal basis set 
$\{ |\omega_0\rangle, |\omega_1\rangle, |\omega_2\rangle \}$ of 
\begin{subequations}
\label{I_Acc_measurement}
\begin{eqnarray}
|\omega_0\rangle 
& = & 
- \sin \frac{\gamma_{\rm Acc}}{2} |1\rangle, 
\\
|\omega_1\rangle 
& = & 
- \frac{1}{\sqrt{2}} |0\rangle 
+ \frac{1}{\sqrt{2}} \cos \frac{\gamma_{\rm Acc}}{2} |1\rangle, 
\\
|\omega_2\rangle 
& = & 
  \frac{1}{\sqrt{2}} |0\rangle 
+ \frac{1}{\sqrt{2}} \cos \frac{\gamma_{\rm Acc}}{2} |1\rangle, 
\end{eqnarray}
\end{subequations}
where $\gamma_{\rm Acc}$ is defined by 
\begin{equation}
\label{gamma_Acc}
\cos \frac{\gamma_{\rm Acc}}{2} = \cot \frac{\pi}{3}.
\end{equation}
This is a typical example of generalized quantum measurement, 
and was demonstrated in laboratory 
for polarization qubit of a photon in Refs. 
\cite{Clarke01b,Mizuno02_ImaxExp}. 
The functional meaning of this quantity $I_{\rm Acc}$ 
is as follows. 
If the receiver applies this detection \textit{separately} 
on each letter states (separable decoding), 
and 
encoding is made such that
each letter states occurs with equal probabilities 
in the set of code words, 
then 
the maximum transmission rate for error free transmission is 
$I_{\rm Acc}=0.5850$ bits/letter.

One may further optimize the quantity 
besides the detection strategy. 
For the ternary set, 
$C_1$ has been carefully studied 
and evaluated to be 0.6454 bits 
\cite{Osaki2000_QCM98_C1,Shor2002}. 
This is attained by discarding one of the three letters 
and using only two of them, 
say $\{ |\psi_0\rangle, |\psi_1\rangle \}$, 
with equal probability $1/2$ 
and 
applying the measurement described by the orthonormal basis, 
\begin{subequations}
\label{C_1_measurement}
\begin{eqnarray}
| \nu_0 \rangle 
& = & 
 \frac{\sqrt{2+\sqrt{3}}}{\sqrt{3}} |\psi_0\rangle 
+\frac{\sqrt{2-\sqrt{3}}}{\sqrt{3}} |\psi_1\rangle, 
\\
| \nu_1 \rangle 
& = & 
 \frac{\sqrt{2-\sqrt{3}}}{\sqrt{3}} |\psi_0\rangle 
+\frac{\sqrt{2+\sqrt{3}}}{\sqrt{3}} |\psi_1\rangle. 
\end{eqnarray}
\end{subequations}
This is schematically illustrated 
in Fig.~\ref{fig:ternary_letter_set}(b).

Now we construct the length two coding. 
For the qubit trine letters, there are nine possible sequences. 
Peres and Wootters showed \cite{PeresWootters91} 
that if one uses only three of them, which are 
\begin{eqnarray} 
\label{ternary_code word} 
|\Psi_{xx}\rangle & = & |\psi_x\rangle \otimes |\psi_x\rangle 
\nonumber\\
	& = & \frac{1}{2}( 1+\cos \phi_x ) |0\rangle|0\rangle 
	  + \frac{1}{2} \sin \phi_x 
	  \Bigl( |0\rangle|1\rangle + |1\rangle|0\rangle \Bigr)
\nonumber\\
	& & + \frac{1}{2}( 1-\cos \phi_x ) |1\rangle|1\rangle, 
\end{eqnarray} 
where $\phi_x = 2 \pi x/3$ $(x = 0,1,2)$, 
as the code word states with equal probability, 
and decodes them
by the square-root measurement defined mathematically by 
\begin{equation} 
\label{SRM} 
| \Pi_{yy} \rangle \equiv 
\left( \sum_x |\Psi_{xx}\rangle 
              \langle\Psi_{xx}| 
\right)^{-\frac{1}{2}}
|\Psi_{yy}\rangle, 
\end{equation} 
then 
$I(X^2:Y^2)=1.3690$ bits of information 
can be retrieved in principle. 
This is larger than twice of $C_1$ $(=0.6454)$. 
The SQCG is 
$I_2/2-C_1=0.0391$ 
which is expected to be accessible in laboratory.

The measurement basis Eq.~(\ref{SRM}) is explicitly written as 
\begin{subequations}
\label{ternary_state_POVM}
\begin{eqnarray}
| \Pi_{00} \rangle 
& = & 
  a \,\, | \Psi_{00} \rangle
+ b \,\, | \Psi_{11} \rangle  
+ b \,\, | \Psi_{22} \rangle, 
\\ [2\jot]
| \Pi_{11} \rangle 
& = & 
  b \,\, | \Psi_{00} \rangle 
+ a \,\, | \Psi_{11} \rangle
+ b \,\, | \Psi_{22} \rangle, 
\\ [2\jot]
| \Pi_{22} \rangle 
& = & 
  b \,\, | \Psi_{00} \rangle 
+ b \,\, | \Psi_{11} \rangle 
+ a \,\, | \Psi_{22} \rangle, 
\end{eqnarray} 
\end{subequations}
where 
\begin{subequations}
\label{cos_g_sin_g}
\begin{eqnarray}
a & = & \frac{4+\sqrt{2}}{3\sqrt{3}} ,
\\
b & = & - \frac{2-\sqrt{2}}{3\sqrt{3}} .
\end{eqnarray}
\end{subequations}
These bases are entangled states 
and 
the measurement described by the POVM 
$\{ \proj{\Pi_{yy}} \}$ 
is a typical example of quantum collective decoding. 
The ternay code word states 
$\{ \ket{\Psi_{xx}} \}$ 
can be described by the real vectors 
in a three dimensional space spanned by 
$\{ |0\rangle|0\rangle, |0\rangle|1\rangle + |1\rangle|0\rangle, 
|1\rangle|1\rangle \}$, 
as seen from Eq.~(\ref{ternary_code word}).  
This ternary set is called the lifted trine, 
and provides interesting insights 
into quantum  measurement problems as discussed by Shor 
\cite{Shor2002,Shor2001}. 
The measurement basis $\{ \ket{\Pi_{yy}} \}$ forms 
another orthonormal basis set in the three dimensional space.  
$\{ \ket{\Psi_{xx}} \}$ and $\{ \ket{\Pi_{yy}} \}$ 
are geometrically depicted in 
Fig.~\ref{fig:ternary_letter_set}(c). 
The code words $\{ \ket{\Psi_{xx}} \}$ are distinguished 
by the projection to each of the nearest $\ket{\Pi_{yy}}$. 
The channel matrix  
$[P(y|x)]=| \langle\Pi_{yy} | \Psi_{xx}\rangle |^2$ 
is expressed as
\begin{eqnarray}
\label{ternary_state_P(j|i)}
[P(y|x)]= \left[
\begin{array}{ccc}
\cos^2 \frac{\gamma}{2} & \frac{1}{2} \sin^2 \frac{\gamma}{2} & 
\frac{1}{2} \sin^2 \frac{\gamma}{2} \\[2\jot] 
\frac{1}{2} \sin^2 \frac{\gamma}{2} & \cos^2 \frac{\gamma}{2} & 
\frac{1}{2} \sin^2 \frac{\gamma}{2} \\[2\jot] 
\frac{1}{2} \sin^2 \frac{\gamma}{2} & 
\frac{1}{2} \sin^2 \frac{\gamma}{2} & 
\cos^2 \frac{\gamma}{2}
\end{array} 
\right] , 
\end{eqnarray} 
where 
\begin{subequations}
\label{cos_g_sin_g}
\begin{eqnarray}
\cos \frac{\gamma}{2} & = & \frac{\sqrt{2}+1}{\sqrt{6}}, \\
\sin \frac{\gamma}{2} & = & \frac{\sqrt{2}-1}{\sqrt{6}}.
\end{eqnarray}
\end{subequations}
In this noiseless model the channel is essentially 
a measurement channel whose ambiguity is due to 
nonorthogonality of the code word states.

While the square root measurement is simply expressed by 
the von Neumann measurement in the three dimensional real space 
in terms of 
Eqs.~(\ref{ternary_state_POVM}), 
this mathematical expression informs us of nothing special 
about physical implementations of the decoder. 
There may be many possible ways to realize effectively 
the measurement channel matrix of 
Eq.~(\ref{ternary_state_P(j|i)}). 
One of systematic and straightforward ways is 
to express the original measurement bases 
as simple separable bases 
plus additional unitary transformation,
and 
to convert the unitary transformation into a quantum circuit 
\cite{Sasaki98_SupAdd,Sasaki98_realization}. 
Along this line, 
we derive a quantum circuit realizing 
$\{ |\Pi_{yy}\rangle \}$.

\begin{figure}
\begin{center}
\includegraphics[width=0.50\textwidth]{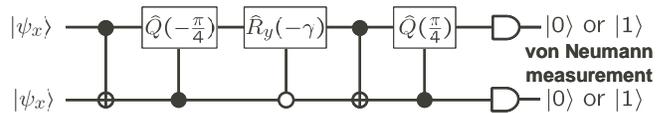}
\end{center}
\caption{\label{fig:two-qubit_circuit}
Quantum circuit to realize the quantum collective decoding 
by the square root measurement $\{|\Pi_{yy}\rangle\}$. 
A received code word state is first transformed 
by the five controlled gates, 
and then is detected 
by a standard von Neumann measurement on each letter separately. 
Nomenclature of the controlled gates is based on 
Ref.~\cite{Barenco95_Gates}.
$\hat{Q}(\varphi)$ is the unitary operator defined as 
$\hat{Q}(\varphi)=\hat{R}_y(\varphi)\hat{\sigma}_z$. 
The open circle notation indicates 
conditioning on the `control' qubit being set to zero. 
}
\end{figure}

Let us rewrite the POVM 
$\{ \proj{\Pi_{yy}} \}$ 
as 
\begin{subequations}
\label{unitary_vonNeumann} 
\begin{eqnarray} 
| \Pi_{00} \rangle & = & \hat{U}^{\dagger} |0\rangle|0\rangle, \\
| \Pi_{11} \rangle & = & \hat{U}^{\dagger} |0\rangle|1\rangle, \\
|   S      \rangle & = & \hat{U}^{\dagger} |1\rangle|0\rangle, \\
| \Pi_{22} \rangle & = & \hat{U}^{\dagger} |1\rangle|1\rangle,
\end{eqnarray}
\end{subequations}
where $|S\rangle = (|0\rangle |1\rangle - |1\rangle |0\rangle)/\sqrt{2}$.  
The unitary operator $\hat{U}$ can be given by 
the matrix representation 
\begin{equation}
\label{unitary_matrix}
\hat{U} = 
\left[
\begin{array}{cccc}
\cos \frac{\gamma}{2} 
  & 0 
    & 0 
      & \sin \frac{\gamma}{2} \\
-\frac{1}{\sqrt{2}} \sin \frac{\gamma}{2} 
  & \frac{1}{2} 
    & \frac{1}{2} 
      & -\frac{1}{\sqrt{2}} \cos \frac{\gamma}{2} \\ [2\jot]
0 
  & \frac{1}{\sqrt{2}} 
    & -\frac{1}{\sqrt{2}} 
      & 0 \\  [2\jot]
-\frac{1}{\sqrt{2}} \sin \frac{\gamma}{2} 
  & -\frac{1}{2} 
    & -\frac{1}{2} 
      & -\frac{1}{\sqrt{2}} \cos \frac{\gamma}{2}
\end{array} 
\right] , 
\end{equation} 
with respect to the separable basis 
$\{
|0\rangle|0\rangle, 
|0\rangle|1\rangle, 
|1\rangle|0\rangle, 
|1\rangle|1\rangle 
\}$. 
Circuit construction for this unitary operator 
can be carried out in the following way. 
With the help of the Gaussian elimination algorithm 
\cite{Reck94_Unitary}, 
$\hat{U}$ can be decomposed into a product of U(2) operators 
$\hat{T}_{[j,i]}$ as
\begin{equation}
\label{U_decomposition}
\hat{U} = \hat{T}_{[2,1]} \hat{T}_{[3,1]} \cdot \cdot \cdot 
\hat{T}_{[4,2]} \hat{T}_{[4,3]}, 
\end{equation}
where 
$\hat{T}_{[j,i]}$ represents 
the two-dimensional rotation operators 
between the {\it i}-th and {\it j}-th basis vectors.
Each operator $\hat{T}_{[j,i]}$ is then converted into 
a quantum circuit 
by using the formulae established by Barenco {\it et al.} 
\cite{Barenco95_Gates}. 
The quantum circuit derived along this line consists of so many 
2 bit basic gates, 
and 
is generally not in the minimal form. 
We further compiled the circuit 
into a much simpler version in a heuristic way. 
The final and possibly the simplest quantum circuit for $\hat{U}$ 
is shown in Fig.~\ref{fig:two-qubit_circuit}. 
It consists of five controlled-unitary gates. 

\section{Implementation}
\label{Implementation}

A favorable qubit trine is made of a flying qubit of photons. 
It would be natural to construct code words 
by the pairs of photons in the same linear polarization states. 
The quantum circuit of Fig.~\ref{fig:two-qubit_circuit} 
then requires photon-photon gates. 
Although the principle of such photon-photon gates have been demonstrated 
experimentally \cite{Turchette95}, 
its precision is still far below the level required 
to access the small superadditive coding gain. 
Even if we rely on other, not flying, qubit systems 
such as trapped ion or molecules in NMR 
for which quantum gating has been demonstrated to date, 
it seems still formidable to run a five-step gating operation 
with the required precision.

Therefore we consider the length two coding 
based on the two physically different kinds of qubit, 
namely, 
a polarization and location qubits of a single photon. 
The first and second letter states of a code word 
are drawn from 
the ternary letter state sets of 
a polarization and a location 
qubits, 
$\{ \ket{\psi_x}_P \}$ and $\{ \ket{\psi_x}_L \}$, 
respectively. 
Then the collective decoding can be realized 
by an optical circuit consisting only of linear passive components, 
and a sufficiently high gating precision can be attained. 
In fact, 
by using the same polarization-location encoding format 
\cite{Takeuchi96,Cerf98_LinearOpt,Spreeuw98}, 
several quantum algorithms 
have been demonstrated experimentally
\cite{Takeuchi00,Kwiat00}.

In the following subsections, 
we first describe the physical implementation 
based on the polarization-location format. 
Then we discuss three kinds of experiments on 
quantum measurement of  
the accessible information, the single--shot capacity, 
and SQCG attained by the second order mutual information. 
These schemes are on a structured scenario of 
the capacity theory as described in Sec.~III. 
They also correspond to the most typical measurements 
for the same qubit trine 
in the framework of quantum measurement theory, i.e. 
the von Neumann measurement, 
the single--shot generalized measurement, 
and the collective measurement.

\subsection{Preparation of optical qubit states}

The polarization qubit consists of 
the horizontal $|0\rangle_P = |H\rangle$ 
and 
the vertical $|1\rangle_P = |V\rangle$ 
polarization states of a single photon 
and prepared by a half waveplate (HWP) 
which acts as 
\begin{subequations}
\label{eq-3-1}
\begin{eqnarray}
| H \rangle 
& \mapsto & 
-\cos 2 \theta |H\rangle + \sin 2 \theta |V\rangle,
\\
| V \rangle 
& \mapsto & 
\sin 2 \theta |H\rangle + \cos 2 \theta |V\rangle,
\end{eqnarray}
\end{subequations}
where 
$\theta$ is the angle of the fast axis from the vertical axis. 
The elements of the ternary set of polarization qubit  
$| \psi_0 \rangle_P$, $| \psi_1 \rangle_P$, $| \psi_2 \rangle_P$ 
can be prepared from the input of 
the $|0\rangle_P$ state 
by setting 
$\theta=0$, 
$\pi/6$, 
$\pi/3$ 
[rad], 
respectively.

The ternary set of location qubit $\{ \ket{\psi_x}_L \}$ 
can be prepared 
by guiding the polarization letter states into two optical paths 
through a polarizing beam splitter (PBS). 
It reflects the vertical polarization 
and transmits the horizontal polarization as 
\begin{subequations}
\label{eq-3-2}
\begin{eqnarray}
| H \rangle_A \otimes |{\rm vacuum}\rangle_B 
& \mapsto & 
        | H \rangle_A \otimes |{\rm vacuum}\rangle_B,
\\
| V \rangle_A \otimes |{\rm vacuum}\rangle_B 
& \mapsto & 
      i | {\rm vacuum} \rangle_A \otimes | V \rangle_B,
\end{eqnarray}
\end{subequations}
where A and B are the labels for the two different optical paths.

\begin{figure}
\begin{center}
\includegraphics[width=0.30\textwidth]{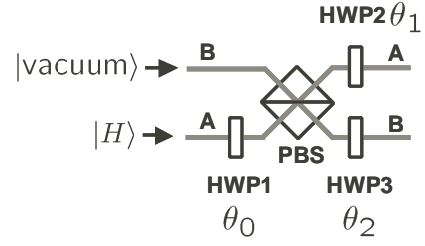}
\end{center}
\caption{\label{fig:optical_circuit_encoding}
Optical circuit for 
polarization-location encoding. 
HWP: half waveplate and 
PBS: polarizing beam splitter. 
}
\end{figure}

The length two coding can be realized 
in the Hilbert space spanned by
the orthonormal bases 
\cite{Takeuchi96,Cerf98_LinearOpt,Spreeuw98}, 
\begin{subequations}
\label{eq-3-3}
\begin{eqnarray}
|00\rangle 
& = & |0\rangle_P \otimes |0\rangle_L 
    = |H\rangle_A \otimes |{\rm vacuum}\rangle_B,
\\
|01\rangle 
& = & |0\rangle_P \otimes |1\rangle_L 
    = |{\rm vacuum}\rangle_A \otimes |H\rangle_B,
\\
|10\rangle 
& = & |1\rangle_P \otimes |0\rangle_L 
    = |V\rangle_A \otimes |{\rm vacuum}\rangle_B,
\\
|11\rangle 
& = & |1\rangle_P \otimes |1\rangle_L 
  = |{\rm vacuum}\rangle_A \otimes |V\rangle_B.
\end{eqnarray}
\end{subequations}
In this space 
the encodings can be performed by 
the optical circuit shown in Fig.~\ref{fig:optical_circuit_encoding} 
which consists of a PBS and three HWPs.
With an input photon initially in the state $|00\rangle$, 
the output of this encoder is given by 
\begin{eqnarray}
\label{eq-3-5}
|\Psi\rangle 
    & = & \cos 2\theta_0 \cos 2\theta_1 |00\rangle
	  - \sin 2\theta_0 \sin 2\theta_2 |01\rangle
\nonumber\\
    & & - \cos 2\theta_0 \sin 2\theta_1 |10\rangle
	  + \sin 2\theta_0 \cos 2\theta_2 |11\rangle,
\nonumber\\
\end{eqnarray}
where $\theta_0, \theta_1$ and $\theta_2$ are 
the angles of the three HWPs.
By controlling these angles appropriately, 
polarization qubit states 
$|\psi_x\rangle_P \otimes |0\rangle_L$, 
location qubit states 
$|0\rangle_P \otimes |\psi_x\rangle_L$, 
and 
the length two code word states 
$|\psi_x\rangle_P \otimes |\psi_x\rangle_L$ 
can be prepared.

Thus in our coding format, 
doubling the transmission resource is realized by 
doubling the {\it spatial resource} 
instead of by using two polarized photons. 
From the viewpoint of communication theory, 
this can be regarded as a kind of pulse position coding 
which is often used 
when the signal power available is severely limited.

\begin{figure}
\begin{center}
\includegraphics[width=0.50\textwidth]{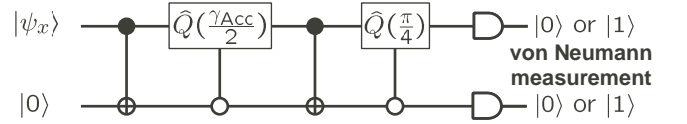}
\end{center}
\caption{\label{fig:I_acc_circuit}
Quantum circuit 
to realize the optimal POVM for the accessible information 
$\{ \proj{\omega_y} \}$ given in Eq.~(\ref{I_Acc_measurement}). 
Nomenclature of the gates is same as in 
Fig.~\ref{fig:two-qubit_circuit}.
}
\end{figure}

\begin{figure}
\begin{center}
\includegraphics[width=0.50\textwidth]{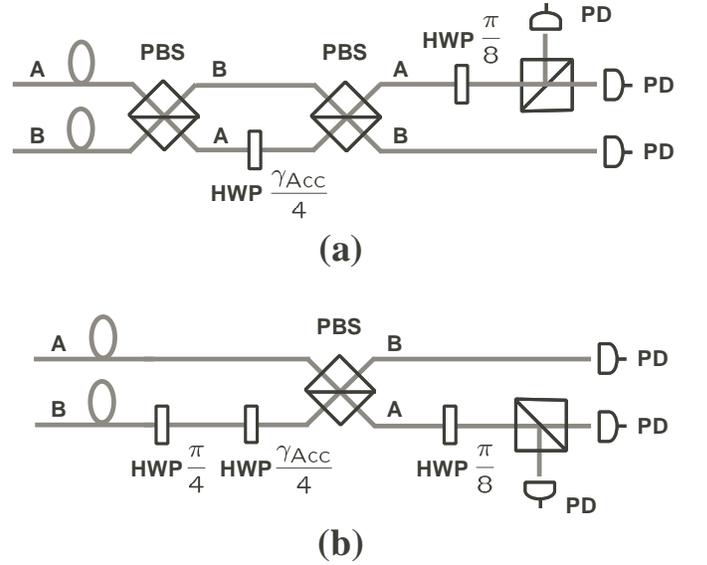}
\end{center}
\caption{\label{fig:optical_circuit_acc}
Optical implementation of the circuit for 
the accessible information in Fig.~\ref{fig:I_acc_circuit} 
for  
(a) polarization qubit and 
(b) location qubit. 
PD: photodetector. 
}
\end{figure}

\subsection{Accessible information}

The POVM for the accessible information $I_{\rm Acc}$ 
generally consists of overcomplete nonorthogonal states, 
which is a typical example of generalized measurement. 
It is well known that such a POVM 
can be implemented by a von Neumann measurement 
in an extended Hilbert space, 
which is called the Naimark extension 
\cite{SasakiBarnettJozsa99}. 
The experiments \cite{Clarke01b,Mizuno02_ImaxExp} 
of such measurement 
were already performed by using polarization qubits. 
We have performed such experiment again for both 
polarization and location qubits. 
We also show the logical quantum circuit of such POVM.

The Naimark extension of the POVM $\{\proj{\omega_y}\}$ given 
in Eq.~(\ref{I_Acc_measurement}) can be implemented by introducing 
an ancillary qubit with that in the initial state $|0\rangle$, 
and by constructing the orthonormal bases 
whose projection onto the original plane 
becomes $\{ |\omega_y \rangle \}$. 
The orthonormal bases are then decomposed into 
the unitary transformation and the von Neumann measurement 
by two qubit separable bases. 
The schematic of this process is shown in 
Fig.~\ref{fig:I_acc_circuit}, 
where the unitary operation is described by 
the quantum circuit consisting of 
four controlled unitary gates. 
This one seems to be almost the simplest circuit. 
As discussed in the previous subsection, 
such quantum circuit can be translated into the optical circuits 
consisting of linear elements. 
Figures~\ref{fig:optical_circuit_acc}(a) and (b) 
show the optical circuits for 
polarization and location qubit signals, 
respectively.

\begin{figure}
\begin{center}
\includegraphics[width=0.50\textwidth]{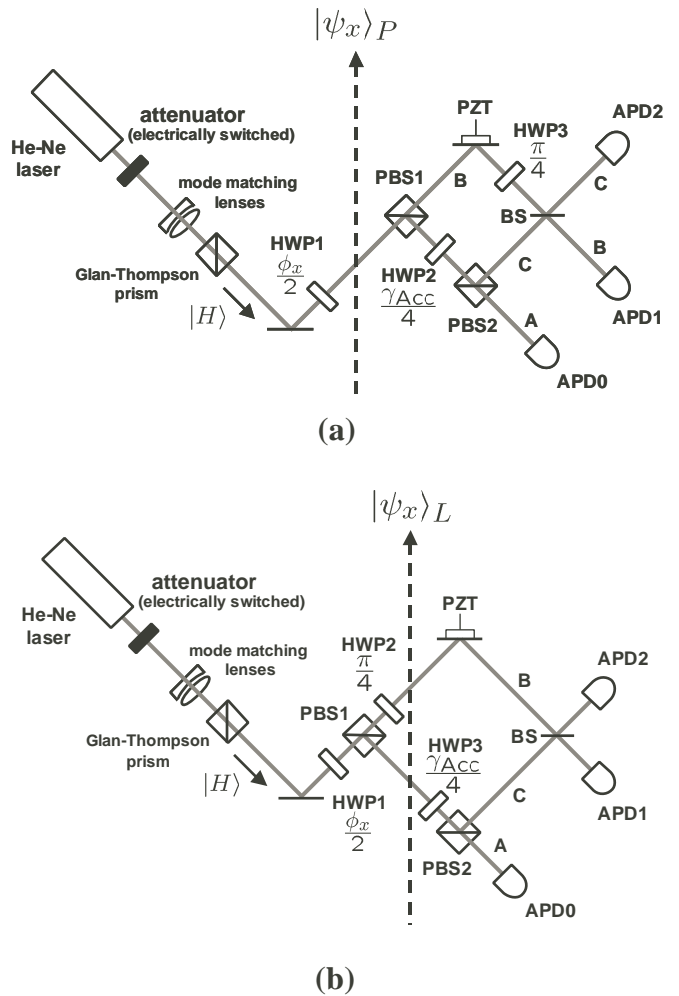}
\end{center}
\caption{\label{fig:optical_circuit_acc_experiment}
Experimental setups for the accessible information for  
(a) polarization qubit and 
(b) location qubit. 
Here $\phi_x=2\pi x/3 + \phi_{\rm off}$ ($x=0,1,2$) 
with the offset angle $\phi_{\rm off}$. 
HWP: half waveplate, PBS: polarizing beam splitter, 
BS: 50:50 beam splitter, and PZT: Piezoelectric transducer. 
}
\end{figure}

The actual experimental setups for 
the polarization and location qubit 
trines are depicted in 
Fig.~\ref{fig:optical_circuit_acc_experiment}(a) and (b), 
respectively. 
In both setups, the left and right of the dashed vertical line 
correspond to the circuits for generating and measuring 
the signals, respectively. 
The signal states are generated 
by varying the angle of HWP1 with $\phi_x=2\pi x/3$ ($x=0,1,2$). 
It should be noted that, 
for practical reason, 
the original circuits in Fig.~\ref{fig:optical_circuit_acc}(a) and (b) 
are modified by using the 50:50 beam splitter (BS) instead of the PBS, 
and the initial state of the ancillary qubit is set 
to the state $|1\rangle$. 

The experimental procedures and techniques are basically the same 
as those in 
Ref.~\cite{Mizuno02_ImaxExp}.
The whole circuit in Fig.~\ref{fig:optical_circuit_acc_experiment} 
consists of a polarization Mach-Zehnder interferometer 
and is controlled by a Piezoelectric transducer (PZT). 
The CW light from a He-Ne laser (Spectra-Physics, model 117A) 
operating at the wavelength of 632.8~nm with 1~mW power 
is strongly attenuated 
by ND filters with a factor of $5 \times 10^{-10}$ 
such that about $10^{-2}$~photons exist on average in the whole circuit.
The attenuated light is purified to 
the horizontally polarized state by a Glan-Thompson prism 
and then injected to the interferometer.
The HWP in the encoder (HWP1) is driven by 
a stepping motor to generate the signal state 
$\{ |\psi_x\rangle_P | x=0,1,2 \}$ 
or 
$\{ |\psi_x\rangle_L | x=0,1,2 \}$ 
sequentially. 
After passing through the circuit, 
the signal photons are guided 
into the silicon avalanche photodiodes 
(EG \& G, SPCM-AQ-141-FC),
APD0--2, 
whose quantum efficiency and darkcount are 
typically 70\% and 100\,counts/sec, respectively, 
through a multimode optical fiber 
with coupling efficiency of about 80\%.
The interferometer is enclosed in a darkened box. 
There are, however, background photons amount to about 
300\,counts/sec, even if no laser light is injected.

The mutual information is evaluated by constructing 
the $3\times3$ channel matrix 
$[ P(y|x) \equiv |\langle\omega_y|\psi_x\rangle|^2 ]$ 
from a statistical data of single photon events 
detected by either of the three APDs 
conditioned on the input state $|\psi_x\rangle$. 
The mutual information thus obtained measures 
the ratio of number of bits retrieved per number of 
total photon counts. 
This event selection allows us to simulate communications of 
sending and detecting photons in {\it pure} state 
one by one through a noiseless channel 
even when a photon source with random arrival times 
\cite{laser}
and 
photodetectors with imperfect efficiencies are used. 
The channel characteristics are then limited 
only by the non-commutativity of the signal states, 
imperfect alignment of the whole interferometer, 
deviation from the lock points, 
and the darkcount of the APDs.

The relative path length of the interferometer is adjusted 
to be a proper operating point 
by using a bright reference beam and the PZT. 
The visibility of the interferometer better than 98\% 
is obtained. 
To circumvent injecting voltage noise from electronics, 
we simply used a low noise voltage source for adjusting and fixing 
mirror positions whereas an electrical feedback system was used 
in Ref.~\cite{Mizuno02_ImaxExp}.
Once the circuit is adjusted, the reference beam is shut off.
The signal light is then guided into the whole circuit.
Photon counts are measured for five-second duration.
This procedure is repeated for each letter state, 
composing a full sequence of measuring the channel matrix.
The temporal stability corresponds to 
the change of the relative path length within 3\,nm 
for at least more than 200\,sec, which causes the error 
in mutual information $\pm$0.005 bits at most.

\begin{figure}
\begin{center}
\includegraphics[width=0.35\textwidth]{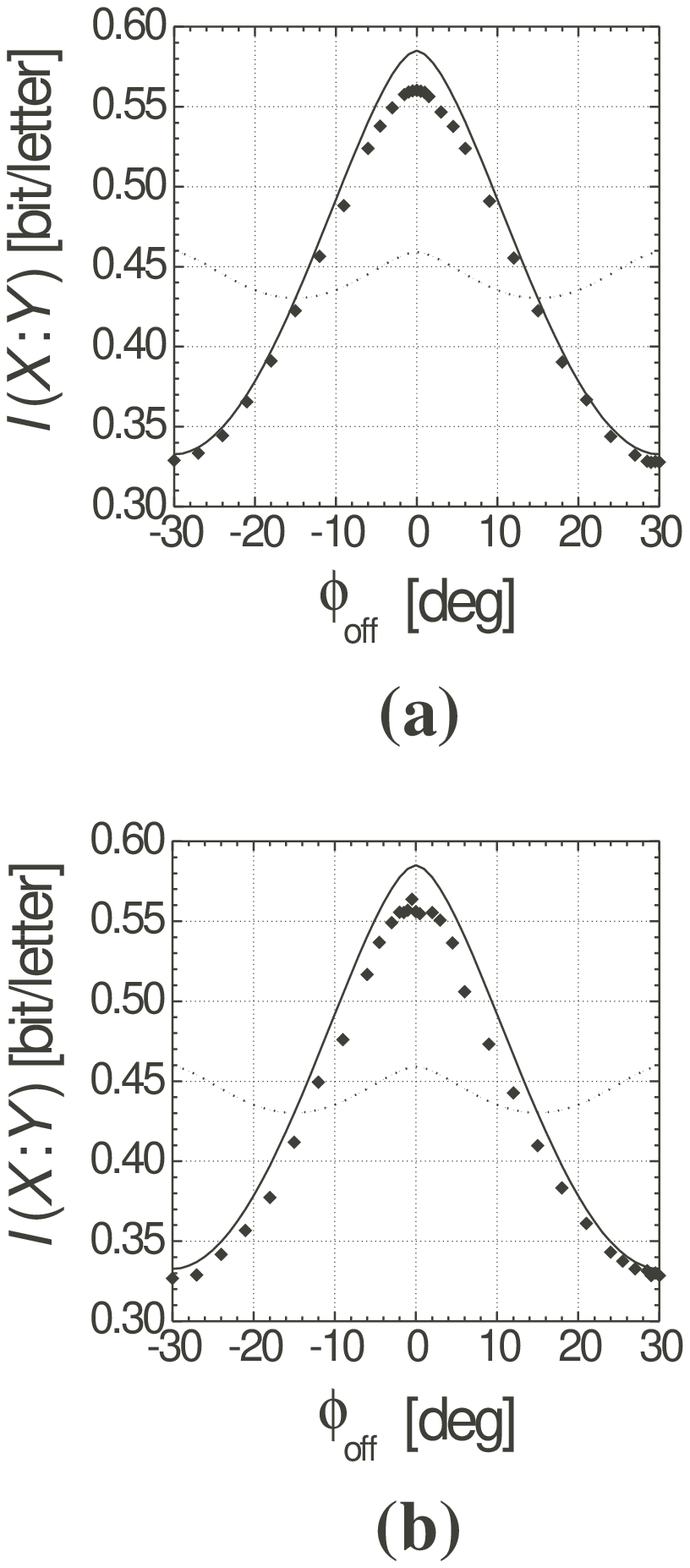}
\end{center}
\caption{\label{fig:mutualinfo_acc}
Measured (filled diamonds) and theoretical (solid curve) 
accessible information for 
(a) the polarization qubit signal and 
(b) the location qubit signal
as a function of the offset angle $\phi_{\rm off}$.
The dashed curve is the mutual information 
obtainable by a standard von Neumann measurement 
\cite{Mizuno02_ImaxExp}.
}
\end{figure}

Figures~\ref{fig:mutualinfo_acc}(a) and (b) 
show the mutual informations measured for 
the polarization and location qubits, respectively. 
The offset angle of the horizontal axis is defined by 
the relative angle $\phi_{\rm off}$ between 
the signal and measurement state sets  
(see Fig.~\ref{fig:optical_circuit_acc_experiment} and its caption). 
The accessible informations is measured at the zero offset angles 
for both the polarization and the location qubits, 
and the results are 
$0.560\pm0.005$~bits/letter 
and 
$0.557\pm0.007$~bits/letter, 
respectively. 
The average visibilities at these points 
are evaluated to be 99.16\% and 99.05\%, respectively.
The difference between the data points and the ideal curve 
is mainly attributed to the imperfection of the PBSs.
The result for the polarization qubit can directly be compared to 
the previous experiments 
\cite{Clarke01b,Mizuno02_ImaxExp} 
and certifies that 
the quality of our interferometer 
is improved from those previous results.

\begin{figure}[t]
\begin{center}
\includegraphics[width=0.50\textwidth]{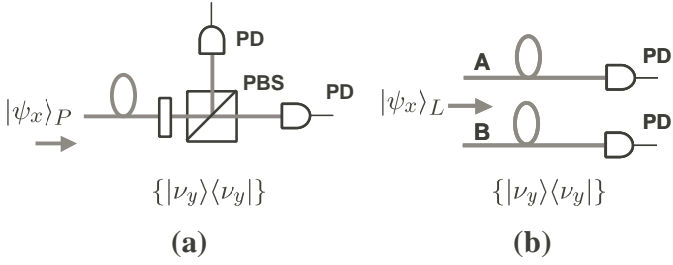}
\end{center}
\caption{\label{fig:C_1_experiment}
Optical circuits for the $C_1$ detection. (a) Polarization qubit, 
(b) location qubit.
}
\end{figure}

\subsection{$C_1$ limit}

The classical limit of the capacity 
$C_1$ $(=0.6454$~bits/letter) is obtained 
by sending only two of three letters with 1/2 probabilities and 
a von Neumann measurement 
in the two-dimensional space 
(Fig.~\ref{fig:ternary_letter_set}(b)). 
The corresponding decoding circuits 
in the polarization and location qubits 
are shown in 
Figs.~\ref{fig:C_1_experiment}(a) and (b).  
For the polarization qubit (Fig.~\ref{fig:C_1_experiment}(a)), 
the polarization of the received photon is rotated by the HWP with 
$\pi/12$ radians, 
and then is discriminated by the PBS followed by APDs. 
Detection of the location qubit is straightforward, 
i.e. 
just detecting a photon at each optical path. 
The measured value is  
$C_1=0.644\pm0.001$~bits/letter, 
for both the polarization and location qubits. 

\begin{figure}
\begin{center}
\includegraphics[width=0.50\textwidth]{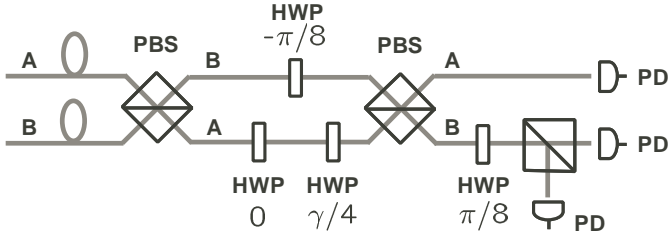}
\end{center}
\caption{\label{fig:optical_circuit_decoding}
Optical circuit for 
the collective decoding 
described by the square root measurement.
}
\end{figure}

\begin{figure*}
\begin{center}
\includegraphics[width=0.70\textwidth]{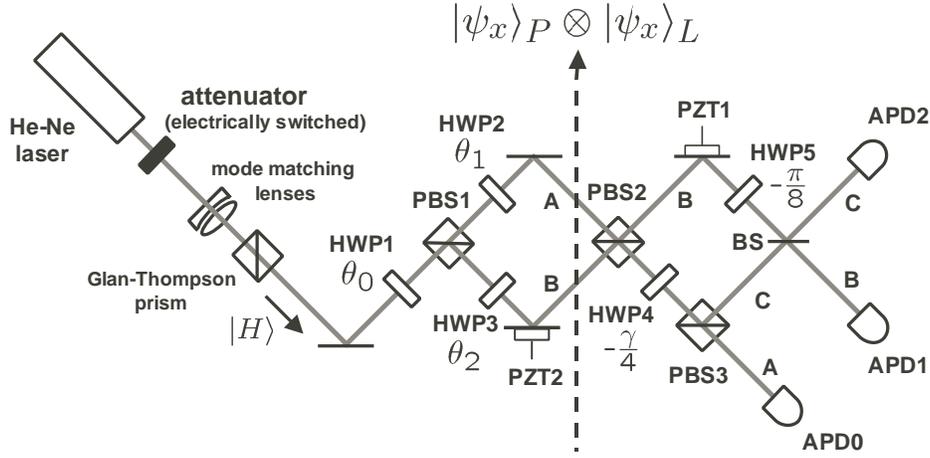}
\end{center}
\caption{\label{fig:experimental_setup}
Experimental setup. 
HWP: half waveplate, 
PBS: polarizing beam splitter, 
BS: 50:50 beam splitter, and PZT: Piezo transducer. 
The HWP angles $\theta_0$--$\theta_2$ are given in the text. 
}
\end{figure*}

\begin{figure} 
\begin{center}
\includegraphics[width=0.40\textwidth]{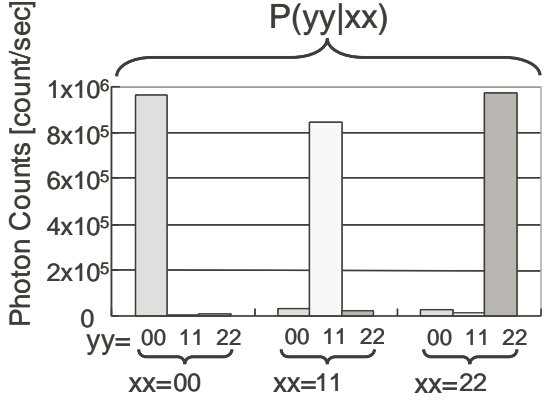}
\end{center}
\caption{\label{fig:channel_matrix}
Histogram of photon counts for the channel matrix elements
$P(yy|xx)=|\langle \Pi_{yy}|\Psi_{xx} \rangle|^2$ (unnormalized).
}
\end{figure}
\begin{figure}
\begin{center}
\includegraphics[width=0.45\textwidth]{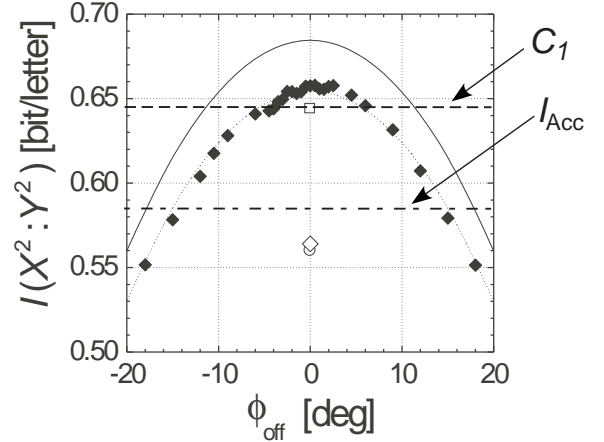}
\end{center}
\caption{\label{fig:mutual_information}
Measured (filled diamonds) and theoretical (solid curve) 
mutual information 
as a function of the offset angle of the code word state set 
$\{|\Psi_{xx}\rangle\}$ 
from the decoder state set 
$\{|\Pi_{yy}\rangle\}$ 
around the vertical axis in Fig.~\ref{fig:ternary_letter_set}(c). 
The dotted curve is just the guide for eyes. 
The experimental and theoretical $C_1$ 
are shown by the square and the dashed line, respectively.
The accessible information $I_{\rm Acc}$ experimentally observed for 
polarization and location qubit, and theoretically predicted are 
shown by the open diamond, open circle, and one-dotted line, respectively.
}
\end{figure}

\begin{figure*}
\begin{center}
\includegraphics[width=0.65\textwidth]{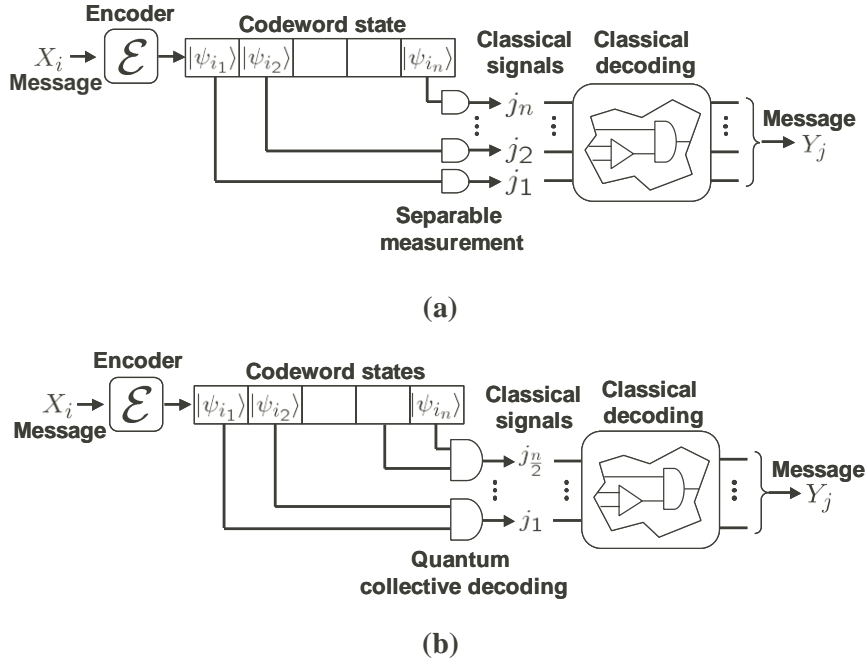}
\end{center}
\caption{\label{fig:ACC_scheme}
(a) Classical channel coding system.
(b) Quantum-classical hybrid channel coding system 
with the letter two quantum collective decoding.
}
\end{figure*}

\subsection{Length two coding}

The length two code word states 
$|\Psi_{xx}\rangle \equiv |\psi_x\rangle_P \otimes |\psi_x\rangle_L$ 
are encoded by the optical circuit of 
Fig.~\ref{fig:optical_circuit_encoding}. 
The angles of HWPs are set as 
\begin{equation}
\label{HWP_angles}
\left\{
\begin{array}{l}
\theta_0 = 
   \frac{1}{2} \arctan 
   \sqrt{ \frac{1-\cos\phi_x}{1+\cos\phi_x} } \\ [1\jot]
\theta_1 = 
   \frac{1}{2} \arctan 
   \left( \frac{-\sin\phi_x}{1+\cos\phi_x} \right) \\ [1\jot]
\theta_2 = 
   \frac{1}{2} \arctan 
   \left( \frac{-\sin\phi_x}{1-\cos\phi_x} \right) 
\end{array}
\right. ,
\end{equation}
where $\phi_x = 2\pi x/3$ ($x=0,1,2$) 
corresponds to the encoding parameter 
in Eq.~(\ref{ternary_code word}).

Figure~\ref{fig:optical_circuit_decoding} shows 
the optical circuit that corresponds to the collective decoding 
circuit in Fig.~\ref{fig:two-qubit_circuit}. 
It is further simplified for practical convenience, 
and the whole experimental setup including encoder 
is shown in Fig.~\ref{fig:experimental_setup}. 
It includes a polarization and a normal interferometers
in Mach-Zehnder arrangements. 
Each interferometer is 
aligned independently to achieve a visibility 
better than 98\%. 
Received code words are decided to be either of $|\Psi_{00}\rangle$, 
$|\Psi_{11}\rangle$, or $|\Psi_{22}\rangle$ according to the reception
of the photon by APD0, APD1, APD2, 
respectively.
Other procedures and conditions are the same as those of the experiment 
for the accessible information extraction.

Figure~\ref{fig:channel_matrix} shows a typical experimental data 
that corresponds to the unnormalized value of each element of 
the channel matrix $[ P(yy|xx) ]$ 
Ideally, the ratio of the diagonal and off-diagonal elements must be 
0.9714 and 0.0143, respectively.
The total events counted for 1 sec is of order 10$^6$, while the average 
count for the off-diagonal elements is about $1.9\times10^4$ 
including the darkcounts of the three APDs. 
The total background count is 2\% of the average count for 
the off-diagonal elements.
The mutual information is evaluated as $I(X^2:Y^2)=1.312 \pm 0.005$~bits. 
The averaged visibility of the whole system 
is evaluated to be 98.48\%, 
which is slightly worse than the result for $I_{\rm Acc}$. 
This degradation is mainly due to 
the relative difference of the polarization axis 
between two interferometers.

For experimental clarity, we measured the variation of 
the mutual information when the code word state set $\{ |\Psi_{xx}\rangle \}$ 
is rotated relative to the decoder state set $\{ |\Pi_{yy}\rangle \}$ around 
the vertical axis in Fig.~\ref{fig:ternary_letter_set}(c).
The rotation is achieved 
by adding an offset angle $\phi_{\rm off}$ to $\phi_x$ 
in Eq.~(\ref{HWP_angles}). 
Figure~\ref{fig:mutual_information} shows the result in which 
the experimental data of the collective decoding (filled diamonds) are 
compared to its ideal curve (solid curve). 
The experimental and ideal values of $C_1$ and $I_{\rm Acc}$ 
are also shown. 
The difference between the data points and the ideal curve 
is attributed again to the imperfection of the PBSs 
and also to the relative difference of the polarization axis 
between two interferometers. 
The obtained mutual information, 0.656$\pm$0.003~bits/letter, 
exceeds the theoretical limit of the classical 
capacity $C_1 = 0.6454$~bits/letter.
Classical length two coding corresponds to the use of 
polarization and location channels at a time 
in a separable decoding circuit construction, which does not 
include any entangling operation.
Then the retrievable information can never exceed $2C_1$.
Our result, therefore, clearly shows the experimental evidence 
of the superadditivity, that is, 
the increase of information more than twice obtained 
by inserting an appropriate quantum circuit to entangle two letter states.

\begin{figure}
\begin{center}
\includegraphics[width=0.45\textwidth]{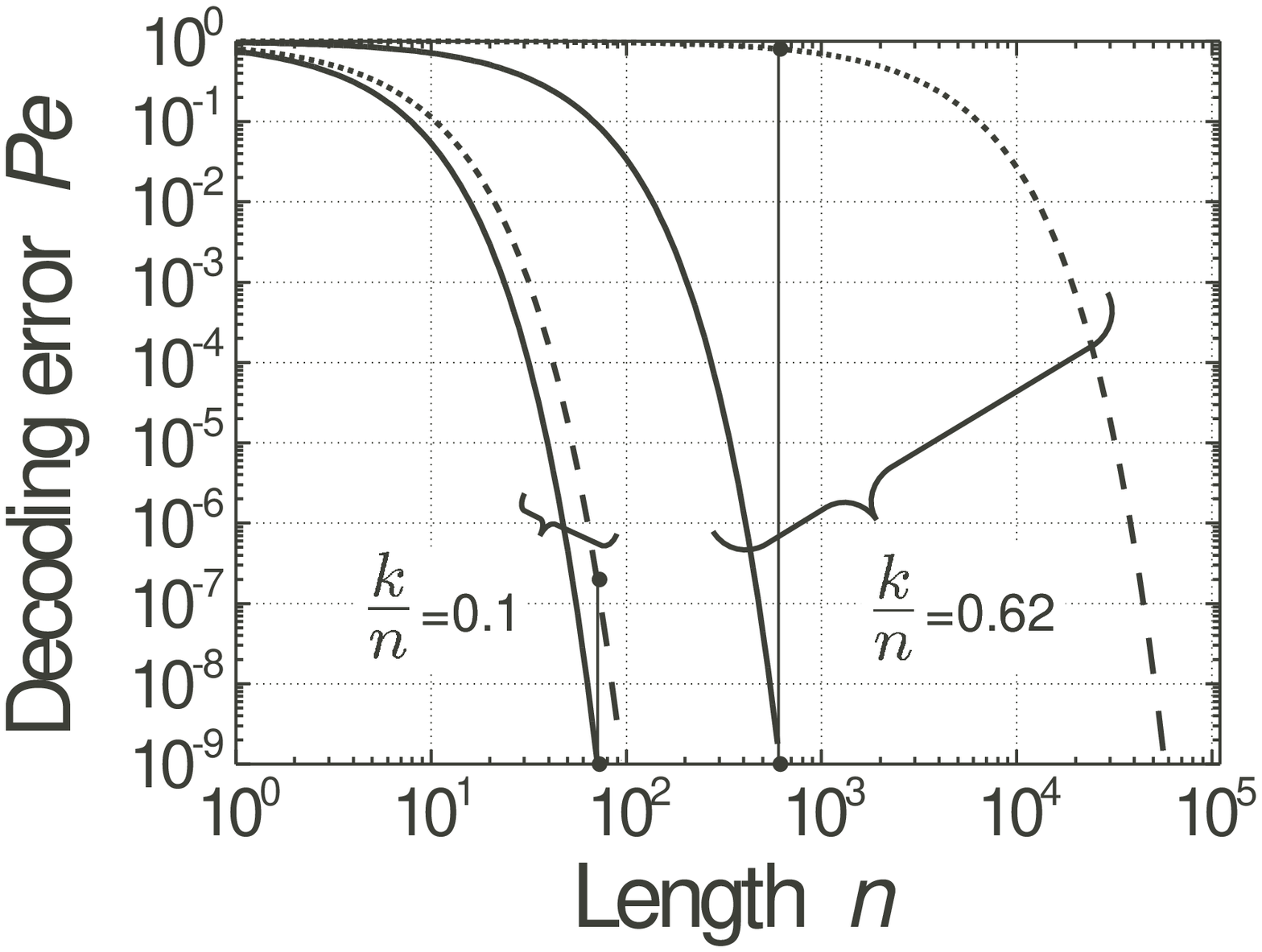}
\end{center}
\caption{\label{fig:Pe-n}
Decoding error probabilities with finite code length 
for the QCHC (solid lines) and the all classical coding (dashed lines). 
}
\end{figure}

\section{Quantum-classical hybrid coding}\label{QCHC}

The SQCG in small blocks is not only 
of proof-of-principle demonstration 
but also 
of practical importance in quantum-limited communications. 
Even two-qubit quantum circuit like 
Fig.~\ref{fig:two-qubit_circuit} 
is useful in boosting the performance of a classical decoder. 
In this section, 
using the case of the ternary letter state set of 
Eq.~(\ref{ternary_state}), 
we show how the two-qubit quantum decoder can be combined with 
a classical decoder to improve 
the total communication performance.

Let us start with a classical channel coding of length~$n$, 
whose schematic is shown in 
Fig.~\ref{fig:ACC_scheme}(a). 
We assume that the letter states are the ternary symmetric states 
given by linear polarizations of a single photon. 
A code word is then physically represented by 
a sequence of optical pulses of single photon, 
which is a tensor product of each letter state. 
At the receiving side, 
each optical pulse enters a photodiode, and is converted into 
an electric pulse, namely, a classical signal. 
This opto-electric conversion is made on each letter state 
separately. 
The classical electric signals are then processed by an electric 
circuit (a classical decoder) to reconstruct an output message.

The capacity attained by this classical coding scheme 
is given by the quantity $C_1$. 
As discussed in Sec.~\ref{Model}, 
$C_1$ (=0.6454 bits/letter) is attained 
by making the code words from only two letter states instead of 
using all three, 
and by performing the binary measurement.
In other words, the channel is equivalent to 
the binary symmetric channel whose error probability is 
characterized by  
\begin{eqnarray}
\label{eq:epsilon}
\epsilon&\equiv& P(1\vert0)=P(0\vert1)
\nonumber\\
&=&{1\over2}
\Bigl( 
1-\sqrt{1 - \vert \amp{\psi_0}{\psi_1} \vert^2} 
\Bigr)
=\frac{2-\sqrt{3}}{4}. 
\end{eqnarray} 
The transmission rate for this channel is defined by $R^{\mathrm C}=k/n$. 
So if $R^{\mathrm C} < C_1=0.6454$, 
one must be able to find codes for which the decoding error approaches 
zero in the limit $n\rightarrow\infty$.

In practice, however, 
one wants to know the decoding error for finite length $n$. 
According to the theory of reliability function 
\cite{Gallager_book}, 
there exists a length $n$ (classical) coding 
attaining $P_{\mathrm e} \le 2^{-n E_r(R)}$, 
where $E_r(R)$ is the lower bound of the reliability function. 
The function $E_r(R)$ is useful to investigate 
the maximum communication performance of channels 
for a given code length.

Let us denote the function $E_r(R)$ for the classical decoding scheme 
as $E_r^{\mathrm C}(R^{\mathrm C})$.
As for the definition of $E_r(R)$ and the expression 
of $E_r^{\mathrm C}(R^{\mathrm C})$, see Appendix. 
We now consider, as examples, two cases where transmission rates are 
low $(R^{\mathrm C}=0.1=0.15\times C_1)$ and high 
$(R^{\mathrm C}=0.62=0.96\times C_1)$. 
In these cases, we obtain $E_r^{\mathrm C}(0.1)=5.218\times10^{-4}$ and 
$E_r^{\mathrm C}(0.62)=0.3150$, respectively, and 
the asymptotic behaviors of the decoding error probabilities are shown in 
Fig.~\ref{fig:Pe-n} (the dashed curves). 
This is a typical error performance obtained by  
averaging over all possible classical codes. 
It means that there must exist at least one code that 
exhibits a performance superior to that shown in Fig.~\ref{fig:Pe-n}.

Now we turn to the other scheme, in which 
the above classical coding is combined with 
quantum channel coding. 
We call such a combination of two coding schemes as 
quantum-classical hybrid coding (QCHC). 
In the following, we show the power of QCHC 
by discussing the QCHC scheme with 
the length two quantum coding. 
Its schematic is shown in Fig.~\ref{fig:ACC_scheme}(b).
Given the length $n$ (assumed to be an even number), 
we consider a classical coding of length $n/2$ 
with the composite letters $\{00, 11, 22\}$. 
In decoding, the received code word state is first processed 
by the two-qubit quantum circuit shown in  
Fig.~\ref{fig:two-qubit_circuit}, 
and then is detected by two photodiodes. 
This is the quantum collective decoding 
consisting of the square root measurement. 
The resulting $n/2$ electric pulses are finally processed by 
a classical decoder.

The channel model of this scheme is equivalent to 
a classical coding of length $n/2$ 
based on the ternary symmetric channel given by 
Eq.~(\ref{ternary_state_P(j|i)}). 
The transmission rate for this channel 
is now defined by $R^{\mathrm{QC}}=(k/n)\log_2 3$  
and it can be raised up to the mutual 
information of this channel, $I(X^2:Y^2)=1.3690$. 
It means that the rate $k/n$ can be raised up to 
$I(X^2:Y^2)/\log_2 3=0.8637$.

Let us denote the function $E_r(R)$ for the QCHC 
as $E_r^{\mathrm{QC}}(R^{\mathrm{QC}})$.
The expression of $E_r^{\mathrm{QC}}(R^{\mathrm{QC}})$ 
is also given in Appendix. 
To compare the performance of $E_r^{\mathrm{QC}}(R^{\mathrm{QC}})$ with 
that of $E_r^{\mathrm C}(R^{\mathrm C})$, 
the rate $k/n$ should be fixed. 
For the same rates as before, $k/n=0.1$ and $k/n=0.62$, 
the transmission rates are given by 
$R^{\mathrm{QC}}=0.1585$ and 
$R^{\mathrm{QC}}=0.9827$, respectively, 
The function $E_r^{\mathrm{QC}}(R^{\mathrm{QC}})$ 
at these rates are evaluated to be 
$E_r^{\mathrm{QC}}(0.1585)=0.8415$ and 
$E_r^{\mathrm{QC}}(0.9827)=9.753\times10^{-2}$, respectively. 
The upper bound of the error probability is then given by 
$P_{\mathrm e}=2^{-(n/2) E_r^{\mathrm{QC}}}$. 
The error probabilities for these examples 
are shown by the solid curves in Fig.~\ref{fig:Pe-n}.

As seen in this figure, at the rate of $k/n=0.62$, 
the decoding error starts to decrease rapidly over $n\sim10000$ 
in the classical coding scheme. 
In the QCHC scheme, 
the error starts to decrease from $n\sim100$, 
and reaches the standard error-free criterion $10^{-9}$ 
around $n\sim600$ (300 composite letter pairs). 
With this code length, 
it is impossible to transmit any information reliably by 
the classical coding scheme 
\cite{CCreliabilityfunction}. 
This improvement can be achieved 
just by inserting the two-qubit quantum 
decoder in front of the classical decoder. 
To achieve the standard error-free criterion 
by classical coding, 
one must use code words of length $n\sim57300$.

As codes get longer, 
the complexity of the decoder, 
such as the total number of arithmetic operations $\chi(n)$, 
increases and eventually limits the effective transmission speed. 
The total decoding time per letter by the device 
with finite speed $\tau_0$ sec/step is given by 
$\frac{\chi(n) }{n}\tau_0$ sec/letter. 
Presumably, this is a limiting factor and thus 
we can define the effective transmission speed as 
$R_\mathrm{eff}=R\frac{n}{\chi(n)\tau_0 }$ bits/sec. 
For some asymptotically good codes, 
the total number of arithmetic operations is typically of order 
$\chi(n)=O((n\log n)^2)$
\cite{Hirasawa80}. 
Therefore, the effective transmission speed behaves as 
$R_\mathrm{eff}\propto\frac{R}{n(\log n)^2}$. 
Then the reduction of code length brought by QCHC will be practically 
significant in the trade-off between performance and complexity. 
In our example, even in the lower rate, the decoding error of the QCHC 
around the standard error free criterion
is two figures smaller than that of the classical coding 
with the same code length.

\section{Concluding remarks}\label{concluding}

In this paper, we have detailed  
the experimental demonstration of 
the superadditive quantum coding gain (SQCG). 
The superadditive quantum coding 
becomes essential 
in the region 
where the noncommutativity of signal states is 
the main ambiguity among signals. 
As a typical example, we have mentioned 
deep space optical communications in Sec.~\ref{introduction}, 
where we have to extract as much information as possible 
from the sequences of heavily attenuated signals. 
Apparently, it is not realistic in such a situation 
to transmit nonclassical states or 
to install quantum repeaters. 
Therefore, only 
quantum decoding can be the core technology 
to achieve the ultimate performance in 
future long-haul optical communications.

Quantum-classical hybrid coding (QCHC) 
is then a promising approach. 
We have shown in Section~\ref{QCHC} that 
even {\it a small scale quantum computing} can enhance 
the effective transmission speed 
when it is used together with large scale classical coding. 
Thus the QCHC allows one to extend 
conventional optical communications technology 
to the quantum limit in a straightforward way. 
This may be contrasted to the fact that 
other known quantum algorithms exhibit the advantage 
over their classical counterparts 
only when a large scale quantum computer is available. 
For the QCHC applications, 
the communication performance of QCHC gets better 
as available scale of quantum computer gets larger.

Finally, we mention the challenges to be overcome 
to bring a QCHC system into reality. 
Although our result clearly demonstrated the principle of SQCG, 
the physical scheme used in our experiment is still not suitable 
for real applications, i.e. is not applicable to 
weak coherent signals. 
To implement a quantum collective decoder for weak coherent signals, 
one must be able to entangle weak coherent optical pulses 
with respect to the degrees of freedom of phase and/or amplitude. 
Heavily attenuated coherent signals can be approximated 
by superpositions of zero and one photon states as 
$|\alpha_k \rangle \approx |0\rangle + \alpha_k |1\rangle$.
It means that the proposals of qubit gating may be applicable 
to our problem.

Quantum gating operation at single photon level has been 
investigated by many authors, but it still remains as a challenging topic. 
One scenario suggested in Ref.~\cite{Buck00_SupAdd} is 
to transfer the information in optical field to 
a multi-level single atom inside a high-finesse optical cavity 
in order to perform gating operations by Raman process. 
Another possible way is to use atomic system as a nonlinear medium  
with the idea of electromagnetically induced transparency
\cite{Paternostro03}. 
Finally, a recent proposal suggests the possibility of 
an optical quantum circuit based on linear optics 
\cite{KnillLaflammeMilburn01}. 
As single photon on-demand sources \cite{Kim99} and 
highly efficient photon detectors become available, 
small scale quantum gating circuits 
for coherent state signals can be realized, 
in principle, with only linear optics.

From the viewpoint of coding theory, on the other hand, 
it is still open to find asymptotically good quantum channel codes. 
In the case of pure state channels, Ref. 
\cite{Hausladen96_capacity} 
tells us that the problem is essentially the selection of 
appropriate sequences for code words, 
and the square root measurement does the decoding. 
It is also important to establish a systematic theory 
to synthesize a quantum circuit 
for intermediate scale collective decoding.

\appendix*
\section{lower bound of the reliability function}
\label{appendix}

In this Appendix, we give the definition 
of reliability function and 
its lower bound, $E_r(R)$. 
The latter one is applied to the channels 
describing the classical coding and the QCHC 
with the length two quantum coding, respectively, 
for the qubit trine signal.

The reliability function is defined as 
\cite{Gallager_book}, 
\begin{equation}
\label{eq:reliability_function_}
E(R) = \lim_{n \to \infty} 
\sup \frac{-\ln P_e (n, R)}{n}, 
\end{equation}
where $P_e(n,R)$ is the minimum error probability 
over all ($n,R$) codes. 
Although the true $E(R)$ for any $R$ has not been clarified yet, 
it is known that its lower bound is given by 
\begin{equation}
\label{eq:reliability_function}
E_r(R) = \max_{\rho} \max_{\{P(x)\}} 
\left[ E_0 (\rho) - \rho R \right] ,
\end{equation}
where 
\begin{equation}
\label{eq:e0}
E_0 (\rho,\,{P(x)}) = - \log \sum_{y} 
\left( \sum_{x} P(x) P(y|x)^{1/(1+\rho)} \right)^{(1+\rho)},
\end{equation}
with $0<\rho\le1$.
The function $E_r(R)$ yields 
the upper bound of an average error probability, $P_{\mathrm e}$, 
for the code with given $n$ and $R$ 
by $P_{\mathrm e} \le 2^{- n E_r (R)}$. 
We also note that, for any symmetric channels, 
$E_r(R)$ is maximized when the all signals are given by 
the equal prior probability distribution \cite{Gallager_book}.

As discussed in Sec.~\ref{QCHC}, 
the channel matrix attaining the capacity 
of the classical coding for the qubit trine signal 
is given by the binary symmetric channel with 
the channel matrix
\begin{equation}
\label{eq:ChannelMatrix_BSC}
[P(y|x)] = \left[
\begin{array}{cc}
   1-\epsilon & \epsilon \\
   \epsilon   & 1-\epsilon 
\end{array}
\right],
\end{equation}
where $\epsilon$ is given in Eq.~(\ref{eq:epsilon}). 
The analytic expression of $E_r(R)$ 
for this channel can easily be derived. 
First we define the quantities, 
\begin{equation}
\epsilon_\rho \equiv
\frac{\epsilon^{1/(1+\rho)}}
{\epsilon^{1/(1+\rho)}+(1-\epsilon)^{1/(1+\rho)}}, 
\end{equation}
and 
\begin{equation}
R_0 \equiv 1-H(\epsilon_1), 
\end{equation}
where 
\begin{equation}
H(\epsilon_\rho) \equiv -\epsilon_\rho\log_2\epsilon_\rho
-(1-\epsilon_\rho)\log_2(1-\epsilon_\rho).  
\end{equation}
After maximizing $E_0(\rho, P(x))$ over $P(x)$ and $\rho$, 
we obtain that 
if $R<R_0$, 
\begin{equation}
E_r^{\mathrm C} (R)=1-2\log_2(\sqrt\epsilon+\sqrt{1-\epsilon})-R, 
\end{equation}
and if $R_0<R<C_1$, 
\begin{equation}
E_r^{\mathrm C} (R)=
\epsilon_{\rho^\ast}
\log_2
\frac{\epsilon_{\rho^\ast}}{\epsilon}
+
(1-\epsilon_{\rho^\ast})
\log_2
\frac{1-\epsilon_{\rho^\ast}}{1-\epsilon}, 
\end{equation}
where ${\rho^\ast}$ is the solution of 
\begin{equation}
R=1-H(\epsilon_\rho). 
\end{equation}

The channel matrix for the QCHC discussed in Sec.~\ref{QCHC} 
is given in Eq.~(\ref{ternary_state_P(j|i)}),  
which is a ternary symmetric channel. 
The expression of $E_r(R)$ for a ternary symmetric channel 
can also be derived with the quantities, 
\begin{equation}
\Gamma_{\rho} = \frac{ \left( \sin^2 \frac{\gamma}{2} 
   \right)^{1/(1+\rho)} }{
   \frac{1}{2} \left( 2 \cos^2 \frac{\gamma}{2} \right)^{1/(1+\rho)} 
   + \left( \sin^2 \frac{\gamma}{2} \right)^{1/(1+\rho)} } ,
\end{equation}
and 
\begin{equation}
R_0 \equiv \log_2 3 - \Gamma_1 - H(\Gamma_1).
\end{equation}
Then if $R<R_0$, 
\begin{equation}
E_r^{\mathrm{QC}}(R) = \log_2 3 
   - 2\log_2 \left( \cos\frac{\gamma}{2}+\sqrt{2}\sin\frac{\gamma}{2} \right),
\end{equation}
and if $R_0<R<I(X^2\,:\,Y^2)$, 
\begin{equation}
E_r^{\mathrm{QC}}(R) = 
\Gamma_{\rho^*} \log_2 \frac{\Gamma_{\rho^*}}{\sin^2 \frac{\gamma}{2}} + 
(1-\Gamma_{\rho^*}) \log_2 \frac{1-\Gamma_{\rho^*}}{\cos^2 \frac{\gamma}{2}},
\end{equation}
where $\rho^*$ is the solution of 
\begin{equation}
R=\log_2 3 - \Gamma_{\rho} - H(\Gamma_{\rho}).
\end{equation}


\begin{references}

\bibitem{Shannon48}
   C.~E.~Shannon,
   Bell System Tech.\ J.\ \textbf{27},
   379 (Part I) and  623 (Part II) (1948). 

\bibitem{Gallager_book}
   R.~G.~Gallager:
   \textit{Information Theory and Reliable Communication}
   (John Wiley and Sons, New York, 1968). 

\bibitem{CoverThomas_book}
   T.~Cover and J.~Thomas:
   \textit{Elements of Information Theory}
   (John Wiley and Sons, New York, 1991). 

\bibitem{Holevo79_QuantCap} 
   A. S. Holevo, 
   Probl. Peredachi Inform. \textbf{15}(4), 3 (1979). 
  (English transl.: Problems of Inform. Transm., 
   \textbf{15}(4), 247 (1980).)

\bibitem{PeresWootters91}	
   A.~Peres and W.~K.~Wootters, 
   Phys.\ Rev.\ Lett.\ \textbf{66}, 1119 (1991).

\bibitem{Sasaki97_SupAdd}
   M.~Sasaki, K.~Kato, M.~Izutsu, and O.~Hirota, 
   Phys.\ Lett.\ A\,\textbf{236}, 1 (1997).

\bibitem{Sasaki98_SupAdd}
   M.~Sasaki, K.~Kato, M.~Izutsu, and O.~Hirota, 
   Phys.\ Rev.\ A\,\textbf{58}, 146 (1998).

\bibitem{Buck00_SupAdd}
   J.~R.~Buck, S.~J.~van~Enk, and C.~A.~Fuchs,  
   Phys.\ Rev.\ A\,\textbf{61}, 032309 (2000). 

\bibitem{Usuda02_SupAdd}
    S.~Usami, T.~S.~Usuda, I.~Takumi, R.~Nakano, and M.~Hata, 
    \textit{Quantum Communication, Computing, and Measurement 3} 
    (Eds. Tombesi, P., and Hirota, O. Kluwer academic/Prenum, New York, 
    2001) 35;   
   	T.~S.~Usuda, S.~Usami, I.~Takumi, and M.~Hata, 
   	Phys.\ Lett.\ A\,\textbf{305}, 125 (2002).

\bibitem{Helstrom_QDET}
    C. W. Helstrom, {\it Quantum Detection and Estimation Theory}
    (Academic Press, New York, 1976).

\bibitem{Gordon62}
   J. P. Gordon, 
   IRE Proc.\,\textbf{50}, 1898 (1962).

\bibitem{Gordon64_bound}
   J. P. Gordon, 
   {\it Quantum Electronics and Coherent Light, 
   Proc. Int. School Phys. ``Enrico Fermi'', Course XXXI}, 
   (Ed. P. A. Miles, New York: Academic Press 1964) 156. 

\bibitem{Lebedev66}
   D. S. Lebdev and L. B. Levitin, 
   Inform.\ Contr.\,\textbf{9}, 1 (1966).

\bibitem {Holevo73_bound} 
   A. S. Holevo, 
   Probl.\ Peredachi\ Inform.\,\textbf{9}(3), 3 (1973). 
   (English transl.: Problems of Inform.\ Transm.,\, 
    \textbf{9}(3), 177 (1973).)

\bibitem{Hausladen96_capacity}
   P.~Hausladen, R.~Jozsa, B.~Schumacher, M.~Westmoreland,
   and W.~K.~Wootters, 
   Phys.\ Rev.\ A\,\textbf{54}, 1869 (1996). 

\bibitem{Schumacher97_capacity}
   B.~Schumacher and  M.~D.~Westmoreland,
   Phys.\ Rev.\ A\,\textbf{56}, 131 (1997). 

\bibitem{Holevo98_capacity}
   A.~S.~Holevo,
   IEEE Trans.\ Inf.\ Theory \textbf{IT-44}, 269 (1998). 

\bibitem{FujiwaraTakeokaMizunoSasaki03_ExpSupAdd}
   M.~Fujiwara, M.~Takeoka, J.~Mizuno, and M.~Sasaki,
   Phys.\ Rev.\ Lett.\,\textbf{90}, 167906 (2003).

\bibitem{Q_cap}
   H.~Barnum, M.~A.~Nielsen, and B.~Schumacher, 
   Phys.\ Rev.\ A\,\textbf{57}, 4153 (1998).

\bibitem{BennettWiesner92}
   C.~H.~Bennett and S.~J.~Wiesner, 
   Phys.\ Rev.\ Lett.\,\textbf{69}, 2881 (1992).

\bibitem{Bennett92}
   C.~H.~Bennett, G.~Brassard,~C.~Crepeau, R.~Jozsa, 
   A.~Peres, and W.~K.~Wootters, 
   Phys.\ Rev.\ Lett.\,\textbf{70}, 1895 (1993).

\bibitem{BennettShorSmolinThapliyal02}
   C.~H.~Bennett, P.~W.~Shor, J.~A.~Smolin, and A.~V.~Thapliyal, 
   IEEE Trans. Inf. Theory \textbf{IT}-48, 2637 (2002). 

\bibitem{FujiwaraNagaoka98}	
   A.~Fujiwara and H.~Nagaoka, 
   IEEE Trans. Inf. Theory \textbf{IT}-44, 1071 (1998). 

\bibitem{Sasaki98_realization}
   M.~Sasaki, T.~Sasaki-Usuda, M.~Izutsu, and O.~Hirota,
   Phys.\ Rev.\ A\,\textbf{58}, 159 (1998).

\bibitem{Davies78}
   E.~B.~Davies,
   IEEE Trans.\ Inf.\ Theory \textbf{IT-24}, 596 (1978).

\bibitem{Levitin95_QCM94}
   L.~B.~Levitin,
   \textit{Quantum Communication, and Measurement}
   (Eds.\ V.~P.~Belavkin, O.~Hirota, and R.~L.~Hudson,
    Prenum, New York, 1995), 439.  

\bibitem{Osaki2000_QCM98_C1}
   M.~Osaki, M.~Ban, and O.~Hirota,
   \textit{Quantum Communication, Computing, and Measurement 2}
   (Eds.\ P.~Kumar, G.~M.~D'Ariano, and O.~Hirota,
    Kluwer academic/Prenum publishers, New York, 2000)
    17.   

\bibitem{SasakiBarnettJozsa99}
   M.~Sasaki, S.M.~Barnett, R.~Jozsa, M.~Osaki, and O.~Hirota,
   Phys.\ Rev.\ A\,\textbf{59}, 3325 (1999).

\bibitem{Clarke01b}
   R.~B.~M.~Clarke, V.~M.~Kendon, A.~Chefles, S.~M.~Barnett, E.~Riis,
   and M.~Sasaki,
   Phys.\ Rev.\ A\,\textbf{64}, 012303 (2001).

\bibitem{Mizuno02_ImaxExp}	
   J.~Mizuno, M.~Fujiwara, M.~Akiba, T.~Kawanishi, S.~M.~Barnett, 
   and M.~Sasaki, 
   Phys.\ Rev.\ A\,\textbf{65}, 012315 (2001). 

\bibitem{Shor2002}
   P.~W.~Shor,
   \texttt{LANL arXiv:quant-ph/0206058}.

\bibitem{Shor2001}
    P.~W.~Shor,
    \textsl{Quantum Communication, Computing, and Measurement 3}
    (Eds.\ O.~Hirota and P.~Tombesi, Kluwer, Dordrecht, 2001) 107.
    Also available at \texttt{LANL arXiv:quant-ph/0009077}.

\bibitem{Reck94_Unitary}
   M. Reck, A. Zeilinger, H. J. Bernstein, and P. Bertani,
   Phys.\ Rev.\ Lett.\,\textbf{73}, 58 (1994).

\bibitem{Barenco95_Gates}	
   A.~Barenco, C.~H.~Bennett, R.~Cleve, D.~P.~DiVincenzo, N.~Margolus, 
   P.~Shor, T.~Sleator, J.~A.~Smolin, and H.~Weinfurter,
   Phys.\ Rev.\ A\,\textbf{52}, 3457 (1995). 

\bibitem{Turchette95}
   Q.~A.~Turchette, C.~J.~Hood, W.~Lange, H.~Mabuchi, and H.~J.~Kimble,
   Phys.\ Rev.\ Lett.\,\textbf{75}, 4710 (1995).

\bibitem{Takeuchi96}
   S. Takeuchi, 
   \textit{Proceedings of the Fourth Workshop on 
   Physics and Computation:
   PhysComp96}, 
   (Ed.\ T.~Toffoli, New England Complex Systems Institute, 
   Boston, 1996), 299.

\bibitem{Cerf98_LinearOpt}	
   N.~J.~Cerf, C.~Adami, and P.~G.~Kwiat, 
   Phys.\ Rev.\ A\,\textbf{57}, R1477 (1998). 

\bibitem{Spreeuw98}
   R.~J.~C. Spreeuw, 
   Found. Phys. 28, 361 (1998);

\bibitem{Takeuchi00}
	S.~Takeuchi,
	Phys.\ Rev.\ A\,\textbf{62}, 032301 (2000).
	
\bibitem{Kwiat00}
	P.~G.~Kwiat, J.~R.~Mitchell, P.~D.~D.~Schwindt, and A.~G.~White,
    J.\ Mod.\ Opt.\,\textbf{47}, 257 (2000).

\bibitem{laser}
We note that the state of a temporal mode at the output of a laser
operating far above its threshold is described as a
Poisson-distributed, mixed photon number state. This description
is further reinforced in our experiment by the use of
photodetectors which project (destructively) the state of the
temporal modes onto number states. Also, nowhere in our experiment
do we determine the phase of the laser, hence a description of the
output in terms of, say, coherent states is inappropriate. Thus,
the light source in our experiment closely approximates a single
photon source, albeit with the photons arriving at random times.

\bibitem{CCreliabilityfunction}
Rigorously speaking, the optimal measurement for $C_1$ may not 
give the maximum $E$ for any transmission rate. 
Nevertheless, it gives almost maximum $E$, especially, 
in the region of high transmission rate 
since the capacity is the maximum transmission rate 
giving a non-negative $E$. 
Therefore, the discussions on 
the advantage of the QCHC is valid. 
To find the optimal measurement which maximizes $E$ is still 
an open problem. 


\bibitem{Hirasawa80}
   S.~Hirasawa, M.~Kasahara, Y.~Sugiyama, and T.~Namekawa, 
   IEEE Trans. Inf. Theory \textbf{IT}-26, 527 (1980). 

\bibitem{Paternostro03}
   M.~Paternostro, M.~S.~Kim, and B.~S.~Ham, 
   Phys.\ Rev.\ A\,\textbf{67}, 023811 (2003).

\bibitem{KnillLaflammeMilburn01}	
   E.~Knill, R.~Laflamme, and G.~J.~Milburn, 
   Nature \textbf{409}, 46 (2001). 

\bibitem{Kim99}
   J.~Kim, O.~Benson, H.~Kan, and Y.~Yamamoto, 
   Nature \textbf{397}, 500 (1999).

\end{references}
\end{document}